\documentclass[]{aa}  

\usepackage{graphicx}
\usepackage{xcolor}
\usepackage{txfonts}
\usepackage{ulem} 

\newcommand{\Ltot}{L_\mathrm{tot}}

\newcommand{\Rvir}{R_\mathrm{vir}}

\newcommand{\LvirGas}{L_\mathrm{gas}^\mathrm{vir}}
\newcommand{\LvirBH}{L_\mathrm{BH}^\mathrm{vir}}
\newcommand{\LvirXRB}{L_\mathrm{xrb}^\mathrm{vir}}
\newcommand{\fgas}{f_\mathrm{gas}}
\newcommand{\Msun}{\rm{M}_{\odot}}
\newcommand{\Lsun}{L_{\odot}}
\newcommand{\lrb}[1]{\left(#1\right)}

\newcommand{\phox}{\textsc{Phox}}

\newcommand{\ulum}{\mathrm{erg\, s^{-1}}}

\begin{document}

   \title{Radial X-ray profiles of simulated galaxies }

   \subtitle{Contributions from hot gas and XRBs}

   \author{S. Vladutescu-Zopp
          \inst{1}\fnmsep\thanks{\email{vladutescu@usm.lmu.de}}
          \and
          V. Biffi\inst{2,3} 
          \and
          K. Dolag\inst{1}\fnmsep\inst{4}
          }
          
    \titlerunning{Galactic X-ray emission in PHOX}
    \authorrunning{Vladutescu-Zopp S., et al.}

   \institute{Universitäts-Sternwarte, Fakultät für Physik, Ludwig-Maximilians-Universität München, Scheinerstr.1, 81679 München, Germany 
             \and
             INAF, Osservatorio Astronomico di Trieste, via Tiepolo 11, I-34131, Trieste, Italy
             \and
             IFPU -- Institute for Fundamental Physics of the Universe, Via Beirut 2, I-34014 Trieste, Italy
             \and
             Max-Planck-Institut f\"ur Astrophysik, Karl-Schwarzschild-Straße 1, 85748 Garching bei M\"unchen, Germany
             }

   \date{Received yy/yy/yyyy; accepted xx/xx/xxxx}

 
  \abstract
   {Theoretical models of structure formation predict the presence of a hot gaseous atmosphere around galaxies. While this hot circum-galactic medium (CGM) has been observationally confirmed through UV absorption lines, the detection of its direct X-ray emission remains scarce. Recent results from the eROSITA collaboration claimed the detection of the CGM out to the virial radius for a stacked sample of MW-mass galaxies.}
   {We investigate theoretical predictions of the intrinsic CGM X-ray surface brightness (SB) using simulated galaxies and connect them to their global properties such as gas temperature, hot gas fraction and stellar mass.}
   {We select a sample of galaxies from the ultra-high resolution ($48\ \rm{cMpc\, h^{-1}}$) cosmological volume of the Magneticum Pathfinder set of hydrodynamical cosmological simulations. We classify them as star-forming (SF) or quiescent (QU) based on their specific star-formation rate. For each galaxy we generate X-ray mock data using the X-ray photon simulator \phox{}, from which we obtain SB profiles out to the virial radius for different X-ray emitting components, namely gas, active galactic nuclei and X-ray binaries (XRBs). We fit a $\beta$-profile to each galaxy and observe trends between its slope and global quantities of the simulated galaxy.}
   {We find marginal differences between the average total SB profile of the CGM in SF and QU galaxies, with the contribution from hot gas being the largest ($>50\%$) at radii $r>0.05\,\Rvir$. The contribution from X-ray binaries (XRBs) equals the gas contribution for small radii and is non-zero for large radii. The galaxy population shows positive correlations between global properties and normalization of the SB profile. The slope of fitted $\beta$-profiles correlates strongly with the total gas luminosity, which in turn shows strong connections to the current accretion rate of the central super-massive black hole (SMBH). All our findings generally agree with the literature. }
   {}


   \maketitle
%

\section{Introduction}
    In the standard cosmological framework, dark matter (DM) halos attract baryonic matter to form galaxies and galaxy clusters. The infalling baryonic matter is then shock-heated to X-ray temperatures ($T\gtrsim 10^6\,K$) in equilibrium with the gravitational potential well \citep{WhiteRees1978, WhiteFrenk1991}, forming a gaseous circum galactic medium (CGM). Due to long cooling times compared to the dynamical time, the CGM is expected to be quasi-static where the majority of cooling processes occur through thermal Bremsstrahlung and line-dominated cooling from different metal species. In addition, there are several feedback mechanisms, such as star-formation and feedback from accreting super-massive black holes (SMBHs) powering active galactic nuclei (AGN), which inject metals and energy into the CGM and cause the CGM to be multi-phase. The interplay between different feedback mechanisms from e.g. AGNs and stellar evolution as well as the refueling of the inner gas reservoir through cooling processes in the CGM play a crucial role in the quenching and growth of galaxies \citep[see review][]{Tumlinson+2017}.
    
    The multi-phase structure of the CGM is apparent from observations. Studies of the absorption and emission lines of hydrogen and metals in the UV band revealed the presence of the warm ($T\sim 10^{5-6}\,K$) phase of the CGM \citep{Tumlinson+2013, Werk+2016, Burchett+2019} for nearby galaxies which provides further evidence that most of the baryonic matter, including metals, is likely bound to the CGM \citep{Stocke+2013, Werk+2014}.
    Direct X-ray emission from the CGM revealed its hot phase ($T\gtrsim 10^6\,K$) in a few nearby massive galaxies \citep{Humphrey+2011,Bogdan+2013,Bogdan+2015,Buote+2017,Li+2017,Das+2019,Das+2020} where the signal to noise (S/N) ratio allowed them to have reliable detection up to $0.15\,R_{200c}$\footnote{In the linear collapse model of structure formation, $R_{200c}$ denotes the radius at which the mean halo density reaches 200 times the critical density}. One of the main challenges in determining properties of the CGM is the low surface brightness (SB) of hot gas at large galacto-centric distances, due to declining gas density. 
    Because the X-ray emissivity of hot plasmas in collisional equilibrium scales with the square of the density, the X-ray SB thus declines faster than the density. In addition, the X-ray foreground of our own milky way (MW) is present in all directions \citep{McCammon+2002} and drowns out most of the signal coming from the low-SB CGM.
    By performing a stacking analysis of survey galaxies, foreground effects can be somewhat mitigated to obtain a statistical signal from the CGM in the soft X-ray band (SXB) \citep{Anderson+2013, Anderson+2015, Li+2018, Comparat+2022, Chadayammuri+2022, Zhang+2024a} which however hampers our ability to quantify other global properties of the CGM such as mass, metallicity and temperature. Recent results from the extended ROentgen Survey with an Imaging Telescope Array (eROSITA) \citep{eROSITA2021} claim the detection of the CGM out to a radius of 300 kpc, close to the virial radius of MW-mass galaxies \citep{Comparat+2022, Chadayammuri+2022, Zhang+2024a}. Both UV and X-ray observations are complemented by analytic models of the temperature and density distribution in the CGM \citep{Faerman+2017, Faerman+2020}.

    Most of the emissivity in the SXB is due to specific metals (e.g. oxygen, neon and iron) which have their transition lines of different ionisation states in the energy range of 0.5-2 keV. Recent studies showed that with high resolution X-ray spectroscopy it may be possible to directly measure the emissivity of specific metal transitions in certain redshift ranges outside of the MW foreground \citep{ATHENA, XRISM, LEM}. This would enable more detailed studies on metal abundances and temperature of the CGM \citep{Wijers+Schaye2022, Truong+2023, Bogdan+2023} and give insights into large-scale anisotropies within the hot X-ray atmosphere  \citep{Truong+2023, Schellenberger+2023, ZuHone+2023}.

    A major component of contamination in X-ray emission are unresolved point sources in the form of X-ray binary (XRB) systems in the stellar field of galaxies. Low-mass XRBs (LMXBs), whose cumulative total luminosity scales linearly with the total stellar mass of the galaxy \citep{Gilfanov2004, Zhang+2012, Lehmer+2016, Lehmer+2019}, are mostly associated with elliptical galaxies with low SFR \citep{Boroson+2011, BogdanGilfanov2011, Lehmer+2020}. High-mass XRBs (HMXBs) are mostly found in galaxies with high SFR \citep{Grimm+2003, Mineo+2012a}.
    From these earlier studies of HMXBs, it is known that both the total number and total luminosity of HMXBs scale linearly with SFR of galaxies. More recent studies found evidence for flatter dependence on SFR for galaxies with low SFR \citep{Kouroumpatzakis+2020, Kyritsis+2024}. Deviations from a linear relation can in principle be connected to a redshift dependence \citep{Lehmer+2016, Aird+2017}, to metallicity \citep{Lehmer+2022} or to stellar age distributions \citep{Lehmer+2017, Gilbertson+2022}. However, the observed flatter relation at low SFR is not consistent with low number sampling of HMXB luminosity functions \citep{Gilfanov+2004b, SVZ:I, Kyritsis+2024}.
    Typically, the total XRB contribution to the total galaxy X-ray luminosity is not exactly known in observations and is modeled empirically for distant galaxies \citep[see e.g.][]{Anderson+2015, Comparat+2022}. Typically, an absorbed power-law spectrum is assumed for unresolved XRB sources. Conservative estimates place the contribution of unresolved XRBs at around $<50\%$ in the SXB \citep{Lehmer+2016, SVZ:I}. This has been recently challenged by \citet{Kyritsis+2024} who estimate the contribution from XRB to be $\sim 80\%$, which is considerably higher compared to previous studies \citep{Mineo+2012b, Mineo+2014, Lehmer+2019, SVZ:I, Riccio+2023}, using stacking results from the eROSITA all-sky survey 1 (eRASS:1).

    This paper aims to shed light on the intrinsic X-ray emission of the CGM in simulated galaxies and connecting it to global properties such as halo temperature, gas fraction and stellar mass. In order to complement results from stacking procedures where information about these global properties is partially lost, we investigate 
    simulated galaxies from the Magneticum Pathfinder suite of hydrodynamic cosmological simulations. We make use of the virtual X-ray photon simulator PHOX \citep{Biffi+2012, Biffi+2018agn} which allows for a self-consistent and detailed multi-component modeling of the X-ray emission coming from various resolution elements of the simulations. In particular, we will be able to make use of this more accurate modeling to account for signals coming from the broad range of temperatures, metallicities and densities present in the hot gas of galaxy-sized halos.
    Additionally, we can directly account for the emission of XRB  \citep{SVZ:I} and SMBH \citep{Biffi+2018agn} sources and give accurate estimates of their contribution. The paper is structured as follows. In section \ref{sec:Magneticum} we briefly highlight details of the cosmological simulation. In section \ref{sec:Phox} we briefly describe the the PHOX algorithm. In section \ref{sec:data_set} we present in detail the retrieval of data from the simulation and how X-ray mock data is created. In addition we give exclusion criteria for a more robust galaxy sample and present our results of X-ray SB profiles on an AGN-cleaned sample in section \ref{sec:Results}. We discuss our results within a broader context in section \ref{sec:Discussion} and summarize our findings in the end.
   
\section{Cosmological hydrodynamical simulation} \label{sec:Magneticum}
    In this work we will make use of the Magneticum Pathfinder Simulations\footnote{Project web page: \texttt{www.magneticum.org}} which are a series of state-of-the-art hydrodynamical cosmological simulations. They explore various cosmological volumes at different resolution levels to understand structure formation and the effect of physical processes on all scales.
    The simulations are performed using an improved version of \textsc{Gadget 3}, which is based on the N-body code \textsc{Gadget 2} \citep{Springel2005}. Fluid dynamics are solved using a Lagrangian prescription for smoothed particle hydrodynamics (SPH). Improvements on the SPH implementation were made on the treatment of artificial viscosity and conductivity \citep{Dolag+2005,Beck+2016}. 
    The evolution of the baryonic component is described through the subgrid implementation of various physical processes. These comprise radiative gas cooling \citep{Wiersma+2009} and heating from a uniform time-dependent ultraviolet background \citep{Haardt+2001} and star formation. The latter is treated as a sub-resolution model with a mass-loading rate proportional to SFR and the inclusion of outflows with wind-velocities of $v_w = 350\,km\,s^{-1}$\citep{SH2003}. The chemical evolution model implemented is the one described in \citet{LT+2004, LT+2007}. The growth and energy feedback from SMBHs adopts the prescription of \citet{Springel2005} and \citet{DiMatteo+2005} with modifications as in \citet{Fabjan+2010}.
    Numerous studies using the Magneticum simulations have been conducted and showed consistency with observations. They reproduce kinematic and morphological properties of galaxies \citep{Teklu+2015,Teklu+2017,Remus+2017a,Schulze+2018,Schulze+2020} as well as chemical properties of galaxies and galaxy clusters \citep{Dolag+2017} and have been employed for studying scaling relations in galaxy clusters \citep{Ragagnin+2019}.
    Furthermore, they present a consistent picture in terms of statistical properties of AGNs \citep{Hirschmann+2014, Steinborn+2016, Biffi+2018agn} and have thus been studied to investigate environmental signatures of AGN activity and star-formation \citep{Rhitarsic+2024}. They have also been successfully used in predicting X-ray properties and signatures of galaxies \citep{SVZ:I, Bogdan+2023} and galaxy clusters \citep{Ragagnin+2022, ZuHone+2023, Churazov+2023, Bahar+2024}.
    
    Similar to \citet{SVZ:I}, we use the same ultra-high resolution run of \textit{Magneticum} called \textit{Box4/uhr}. It represents a $(48\,\rm{h^{-1}\,cMpc})^3$ co-moving volume with a mass resolution of $m_{\mathrm{DM}} = 3.6\times10^7\,\Msun$ and $m_{\mathrm{gas}} = 7.3\times10^6\,\Msun$ for dark matter and gas respectively corresponding to $576^3$ particles. The simulation adopts the cosmological parameters from the \textit{Wilkinson Microwave Anisotropy Probe} (\textit{WMAP7}) \citep{WMAP7} to model initial conditions ($h=0.704$, $\Omega_M = 0.272$, $\Omega_{\Lambda} = 0.728$, $\Omega_b=0.0451$, $\sigma_8 = 0.809$).   
   
\section{PHOX X-ray photon simulator} \label{sec:Phox}
    In this section we describe the general algorithm which we used to obtain synthetic X-ray spectra of the baryonic component in the simulations. In particular, we employ the \phox{} algorithm \citep{Biffi+2012, Biffi+2013} to produce photons according to the intrinsic properties of the simulation.
    The X-ray photon simulator \phox{} operates on three separate modules, summarized as follows.
    
    The first module (\textsc{unit 1}) is responsible for the conversion of hydrodynamical simulation input into a discrete photon distribution. This is done by considering each possible source in the simulation and computing an idealized spectrum.
    In the case of a gaseous source, a single temperature APEC thermal emission model \citep{APEC} is assumed \citep{Biffi+2012, Biffi+2013}, 
    which directly depends on intrinsic properties such as temperature, density and total metallicity or variable chemical abundances.
    In case of a SMBH source we assume an intrinsically absorbed powerlaw spectrum with variable slope and column density which mimics torus absorption \citep{Biffi+2018agn}. This allows for all SMBHs to become a potential AGN.
    In case of a stellar source we assume that the underlying stellar population hosts a XRB component which is either HMXBs or LMXBs. Each seeded XRB also gets assigned an absorbed powerlaw spectrum with fixed slope ($\Gamma_{\rm{LMXB}}=1.7$ for LMXBs, and $\Gamma_{\rm{HMXB}}=2$ for HMXBs) and fixed column density ($N_H^\mathrm{xrb}=2\cdot 10^{21}\, \mathrm{cm}^{-2}$) \citep{SVZ:I}.
    Spectral computations make heavy use of the XSPEC\footnote{v12.12} library interface \citep{XSPEC}. Photons are then sampled stochastically from the computed model spectra for each component assuming a fiducial exposure time and collecting area.
        
    In the second unit (\textsc{unit 2}), the photon data generated by the first module is projected along a random direction through the simulation box and stored in photon lists. The projection accounts for redshift effects on photon energies induced by Doppler shifts from the line-of-sight velocity component of the emitting source as well as angular and luminosity distance induced redshift corrections. 
    Additionally we assume a weak galactic foreground absorption component with column density $N_H^{gal} \sim 10^{20}$ cm$^{-2}$. In all cases, the absorption follows the \texttt{TBABS} model \citep{TBABS} as implemented in XSPEC
    Additionally, a spatial selection can be considered in order to process only small sub-volumes of the photon data.
        
    As a last step (\textsc{unit 3}), the photon lists produced by the second module can be convolved with response matrices of designated X-ray telescopes or can be directly processed by existing X-ray telescope simulators such as SIXTE\footnote{https://www.sternwarte.uni-erlangen.de/sixte/} \citep{Dauser+2019}.

    This approach has been previously applied to study properties of the inter-cluster medium (ICM) of simulated galaxy clusters \citep{Biffi+2012, Biffi+2013,Biffi+2014,Biffi+2015,Cui+2016}, the contamination of ICM emission by AGN \citep{Biffi+2018agn} and to study the contribution of XRB emission in galactic X-ray scaling relations and spectra for simulations of galaxies \citep{SVZ:I}. The \phox{} code is constructed in a general manner such that it can be expanded easily to include the treatment of various X-ray sources which can be constrained from source properties tracked by the simulations.

\begin{figure*}
        \centering
        \includegraphics[width=1\linewidth]{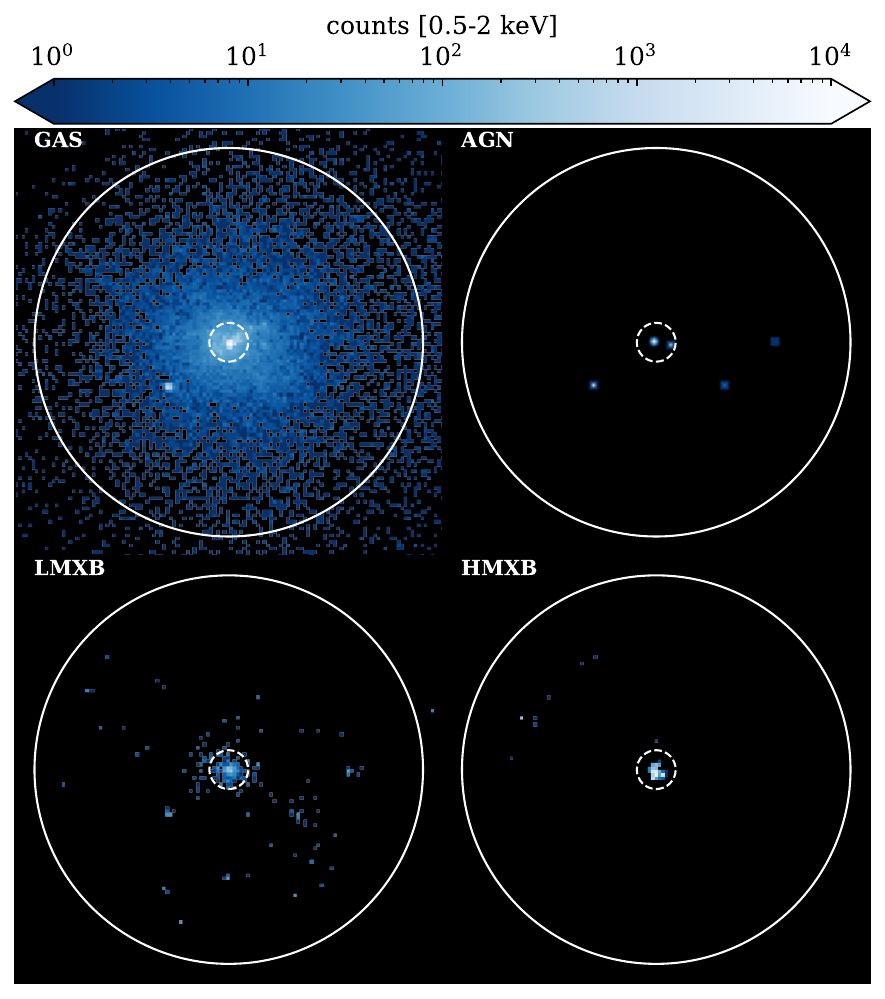}
        \caption{X-ray mock images of the poster-child star-forming disk galaxy from \citet{SVZ:I}. The fiducial orientation of its stellar component is face-on. The outer solid white circle indicates the virial radius and the inner dashed circle indicates 10\% of the virial radius. The color indicates the total photon in the SXB per pixel.}
        \label{fig:sid13633}
    \end{figure*}

    In figure \ref{fig:sid13633} we show the result of running \textsc{unit 1} and \textsc{unit 2} on a face-on disc galaxy from our sample, identified as a ``poster-child'' star-forming disk galaxy in \citet{SVZ:I}. The white outer solid circle corresponds to the virial radius of the galaxy while the inner white dashed circle corresponds to 10\% of the virial radius. The colorbar indicates the photon counts per pixel within the energy range of 0.5-2 keV, assuming a fiducial exposure of 1 Ms and effective area 1000 $\rm{cm^2}$. Black pixels have no photon counts. Each panel shows the same field of view of the galaxy with a different X-ray emitting component which is indicated in the top left corner of the panel.

\section{Simulated data set}  \label{sec:data_set}
    From the selected cosmological volume described in section \ref{sec:Magneticum} we extract 1319 galaxy-sized halos at redshift $z=0.0663$ which roughly corresponds to an angular diameter distance of $\mathcal{D}_A \approx 260$ Mpc. They were selected using the \textsc{subfind} algorithm \citep{subfind2001, subfind2009} which defines halos according to a density threshold. We only considered the main central sub-halo within each parent-halo. We further require them to have stellar mass $10^{10}\Msun < M_* < 10^{12}\Msun$ within a sphere of the virial radius $\Rvir$ around the halo center. The lower mass cut accounts for resolution limitations, 
    the higher mass cut excludes the most massive group-sized halos.
    Resolution limitations arise for halos in which the total number of particles is of the order of the minimum number required for the SPH interpolation, which makes hydrodynamical quantities unreliable. 
    Our initial selection is the same as in \citet{SVZ:I} and consists of 324 star-forming (SF) and 995 quiescent (QU) galaxies. The classification into SF and QU galaxies is based on the specific star-formation rate ($\rm{sSFR}=\rm{SFR}/M_*$), where SF galaxies have $\log \mathrm{sSFR}>-11$. For each galaxy we derive SFRs from the stellar mass born in the past 100 Myrs of the simulation, in line with typical estimators from the literature for X-ray studies of galaxies \citep{Mineo+2012a, Mineo+2012b, Lehmer+2016, Lehmer+2019, Kouroumpatzakis+2020}. 
    
    Following our initial selection, we obtain X-ray photons for each galaxy by first applying \textsc{unit 1} with fiducial exposure time of $T_\mathrm{exp} = 2$ Ms and effective area $A_\mathrm{eff} = 1000$ cm$^2$ on the full simulation volume. For the gaseous component, the idealized spectrum follows an APEC model scaled with the total metallicity of each gas source.
    In contrast to \citet{SVZ:I}, we however did not impose an intrinsic ISM-absorption component on the gas emission for star-forming galaxies in addition to the global foreground Galactic absorption (see further discussion in section \ref{sec:Discussion}).
    Emission from SMBH and XRB follow the modeling by \citet{Biffi+2018agn} and \citet{SVZ:I} respectively.
    
    Next, we apply \textsc{unit 2} on each selected galaxy by projecting the produced photons within a cylindrical volume with base radius of $\Rvir$ and depth of $2\cdot\Rvir$ around the galaxy center. The galaxy center is the position of the most-bound particle according to \textsc{subfind}. The chosen l.o.s.\ is parallel to the $z$-axis $\pmb{\hat{e_z}}$ of the underlying simulated volume. We chose $\Rvir$ as a scale-free radius to have a sufficient representation of the gravitationally bound region. In our analysis, we will not include an instrumental response because we opted to predict intrinsic properties from the simulations.
    
    From the projected photons we construct surface brightness (SB) profiles by radially binning photons in the plane perpendicular to the l.o.s.\  centered on the minimum of the gravitational potential. For each radial bin, we take the sum of all photon energies in a chosen energy range and normalize by the area of the annulus. We thus obtain 
    \begin{equation}
        \label{eq:SB}
        S_X\left(R_{i+\frac{1}{2}}\right) = \dfrac{\sum\limits_{j,\,R_i<r_j<R_{i+1}} \hspace{-10pt}\epsilon_j}{A_\mathrm{eff} T_\mathrm{exp}\,\pi\left(R_{i+1}^2-R_i^2\right)}\times 4\pi\mathcal{D}_L^2\,,
    \end{equation}
    for the SB profile, with $\mathcal{D}_L$ being the luminosity distance, $\epsilon_j$ and $r_j$ the photon energy and its projected radial distance from the center, $R_i$ the edges of the radial bins and $i+\frac{1}{2}$ assigns the determined SB value to the center of the radial bin. The same construction applies to all considered X-ray components. 
    
    Throughout this work, luminosities are given in the rest-frame in the energy range of 0.5-2 keV, if not otherwise noted.
    For the total luminosity $\Ltot$ of each galaxy we integrate SB profiles up to $\Rvir$ and take the sum of each component. 
    By construction, the X-ray emission of each galaxy is derived from the photons emitted by all the X-ray sources (gas, SMBHs, XRBs) within the galaxy $R_{\rm vir}$, with no distinction between central galaxy and substructures.
    
    \subsection{Sample cleaning}\label{sec:AGN-cleaning}
\begin{figure}
        \centering
        \includegraphics[width=\hsize]{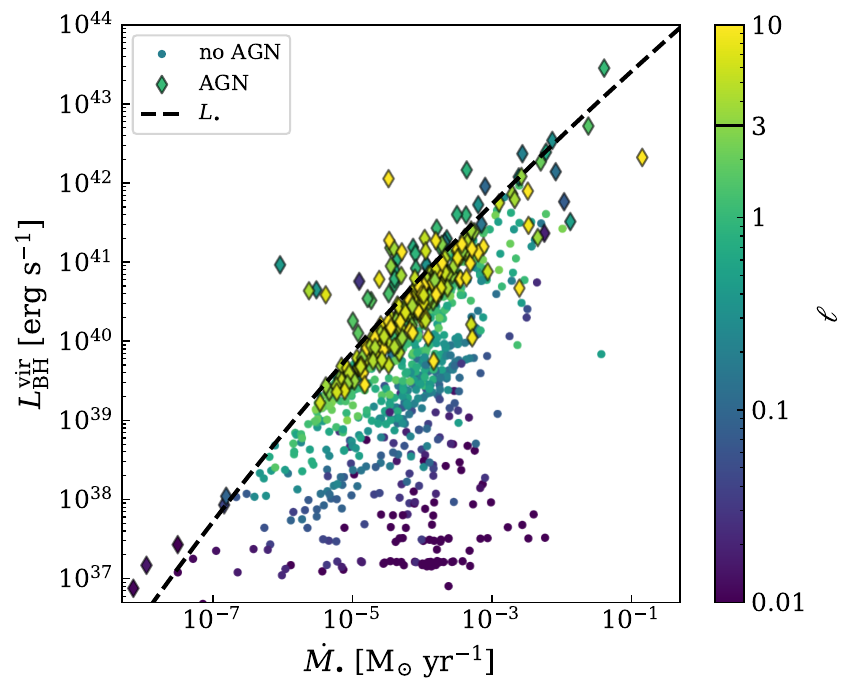}
        \caption{
            Integrated luminosity of SMBH sources ($L_\mathrm{BH}^\mathrm{vir}$) for each galaxy in the 0.5-2 keV energy band as a function of the accretion rate of the central SMBH ($\dot{M}_\bullet$). The dashed line indicates the upper limit set by the bolometric luminosity of the central SMBH (see criterion 3). Data points above the dashed curve host more than one X-ray bright SMBH source within $\Rvir$. Each galaxy is color coded by the ratio $\ell$ (eq. \eqref{eq:agn_ratio}) between the integrated luminosity of SMBH sources and the other two components (gas + XRB). Galaxies which have been excluded following our AGN classification scheme are highlighted by thick edges and have been removed for the final sample.
        }
        \label{fig:agn_exclusion}
    \end{figure}

    After determining the luminosity of each component, we further clean our initial galaxy sample by applying exclusion criteria for AGN galaxies which we partially adapt from \citet{Lehmer+2016}. In this way, we can focus our investigation on a well behaved subsample of galaxies, in which there is no dominant contamination from AGN emission. This will provide us with a more solid base to interpret the behavior of SB profiles in typical galaxies. In particular we use the following criteria: 
    \begin{enumerate}
        \item If the total luminosity of a galaxy is
        \begin{equation}
            L_\mathrm{tot}^{0.5-7\,\mathrm{keV}} > 3\cdot10^{42}\, \mathrm{erg\, s^{-1}}\, ,
        \end{equation} in the 0.5-7 keV energy band, we consider the source to be an AGN. This is directly taken from \citet{Lehmer+2016} and was also recently employed by \citet{Riccio+2023} as an exclusion proxy.
        \item If the integrated luminosity ratio $\ell$ between SMBH sources ($L_\mathrm{BH}^\mathrm{vir}$) and the other components ($L_\mathrm{gas}^\mathrm{vir}$ and $L_\mathrm{xrb}^\mathrm{vir}$ respectively) within the virial radius is
        \begin{equation} \label{eq:agn_ratio}
            \ell = \dfrac{\LvirBH}{\LvirXRB+\LvirGas}>3\, ,
        \end{equation}
        we also consider the source to be an AGN. This second condition was inspired by the requirement to fulfill a pure Lx-SFR scaling relation in \citet{Lehmer+2016}. They allowed the total X-rayluminosity to be three times larger than the SFR scaling relation from \citet{Alexander+2005} which is based on the radio luminosity at 1.9 GHz. The argument is that at this frequency, radio emission should mostly come from star-formation while any excess would be associated to an AGN. Without appropriate tracers for radio emission from the simulation however, we reformulate their criterion to be instead the ratio between the X-ray power from SMBH sources and the combined X-ray power of gas and XRBs. With this we can exclude galaxies which are clearly dominated by an AGN.
        \item If $L_\mathrm{BH}^\mathrm{vir}$ exceeds the expected luminosity of the central SMBH ($L_\bullet$, see below for definition) 
        \begin{equation} \label{eq:agn_upper_limit}
            \log L_\mathrm{BH}^\mathrm{vir} > \log L_\bullet \, ,
        \end{equation}
        we remove the source from our sample. By construction, each SMBH source gets assigned its spectral parameters dependent on its bolometric luminosity such that
        \begin{align}
            \log L_\bullet \leq &\log L_\mathrm{bol} + \sigma \nonumber\\
            &-1.65 - 0.22\mathcal{L} - 0.012\mathcal{L}^2 + 0.0015\mathcal{L}^3\, ,
        \end{align}
        where $\mathcal{L} = \lrb{\log\frac{L_\mathrm{bol}}{\Lsun}-12}$ \citep[see][]{Marconi+2004, Biffi+2018agn}.
        The second term $\sigma = 0.1$ denotes the maximum of randomized uniform noise which was added to the bolometric correction by \citet{Biffi+2018agn}. The latter four terms denote the bolometric correction for the soft X-ray band (0.5-2 keV). The bolometric luminosity ($L_\mathrm{bol}$) of the central SMBH is $L_\mathrm{bol} = \varepsilon_r\dot{M}_\bullet c^2$, where $\dot{M}_\bullet$ is the accretion rate of the central SMBH, $\varepsilon_r=0.1$ is the radiative efficiency, $c$ is the speed of light.
        In our modeling, this acts as an upper limit for the luminosity of a single SMBH source, because we employ an effective torus model with intrinsic absorption for SMBH sources which effectively lowers the power output in the soft band.
        In general, multiple emitting SMBH sources can be present in our simulated galaxies. Thus, the criterion expressed by eq. \eqref{eq:agn_upper_limit}) is equivalent to excluding systems hosting more than one luminous SMBH source within $\Rvir$.
    \end{enumerate}
    In figure \ref{fig:agn_exclusion} we show $L_\mathrm{BH}^\mathrm{vir}$ of the projected volume against $\dot{M}_\bullet$ of the central SMBH for the complete sample of galaxies. Colors indicate the ratio $\ell$ (eq. \eqref{eq:agn_ratio}). Galaxies that were classified as an AGN according to our exclusion criteria are marked as diamonds.
    The dashed diagonal line indicates the upper limit which we impose for the collective emission of SMBH sources (see eq. \eqref{eq:agn_upper_limit}).
    From the total sample, 86 SF and 128 QU galaxies are classified as an AGN according to criterion 1.
    According to criterion 2, 11 SF and 1 QU galaxies have large $\ell$ and are mostly found for central accretion rates $10^{-5}\,\Msun\,\mathrm{yr}^{-1}<\dot{M}_\bullet<10^{-3}\,\Msun\,\mathrm{yr}^{-1}$.
    With criterion 3 we find 31 SF and 21 QU galaxies for which the integrated luminosity $L_\mathrm{BH}^\mathrm{vir}$ is above the upper limit set by the central SMBH accretion and thus host more than one bright SMBH source within $\Rvir$.
    Accounting for overlaps, all criteria together thus reduce the full sample to 338 SF and 727 QU normal galaxies.

\subsection{Determination of galaxy properties}
\label{sec:Determination}
In this section we outline the direct estimate of galactic properties from the simulation. In contrast to the X-ray data retrieval, we do not limit ourselves to 2D projected quantities but make full use of the 3D information available from the simulation. We first select all resolution elements within a sphere of $\Rvir$ around each galaxy center. Then we filter for all particles which are gravitationally bound to the central halo according to the \textsc{subfind} identification.
This procedure allows us to remove the substructures for each considered system from our analysis, and derive the properties of the central galaxies only.
For the stellar mass $M_*$, we take the sum of the mass of each stellar particle in the matched list. For the hot gas fraction $\fgas$ and the halo temperature $k_B T$, we first select gas particles from the matched list that are considered X-ray emitting. Specifically, we select those that are not star-forming and not multiphase (do not represent cold gas), that have a temperature $10^5\,K \leq T\leq5.85\cdot 10^8\,K$ and an intrinsic density of $\rho<5\cdot10^{-25}\,\rm{g}\,\rm{cm}^{-3}$.
To obtain $\fgas$, we take the ratio between the summed mass of X-ray emitting gas particles and all gravitationally bound particles (including stars, dark matter and gas). To derive a single halo temperature, we calculate the emissivity weighted average of the selected gas particles. This approach will yield values close to a spectral temperature.
The emissivity weights were calculated directly from the Astrophysical Plasma Emission Database (APED)\footnote{http://www.atomdb.org} tables used by APEC~\citep{APEC} accounting for individual metal abundances assuming solar abundance from \citet{AnGr1989}.

\section{Galaxy X-ray surface brightness} \label{sec:Results}
    \begin{figure*}
        \centering
        \includegraphics[width=\hsize]{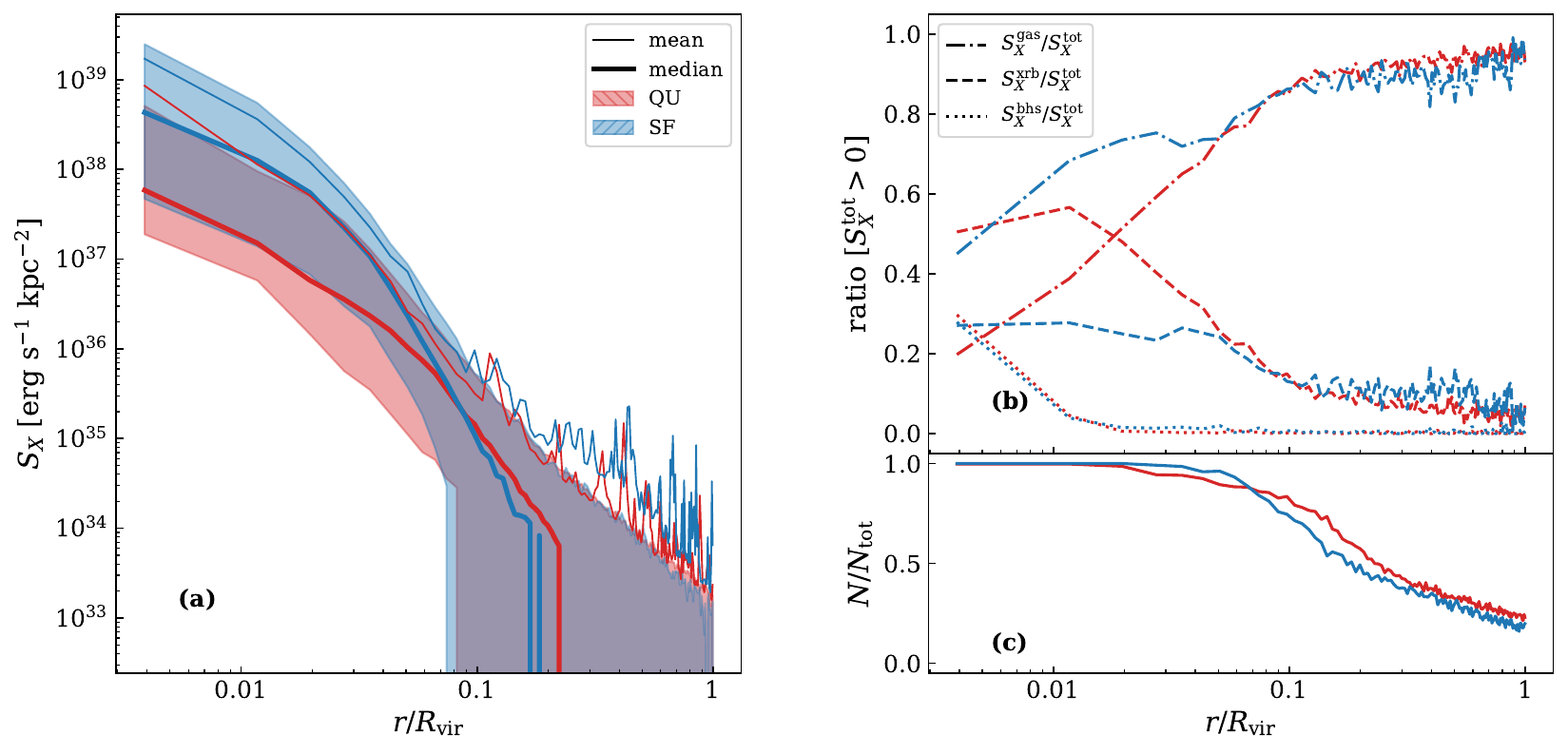}
        \caption{
             \textit{(a)}: Average SB profiles (blue for SF, red for QU) of the normal galaxy sample in the 0.5-2 keV energy band. Thin solid lines indicate the mean total SB. Thick lines indicate the median total SB. The shaded area around the thick lines corresponds to the 16-84 percentile ranges. \textit{(b)}: Mean ratio of the SB profiles of one component (gas: dash-dotted; XRB: dashed; SMBH: dotted) towards the total SB. At each radius we determine the ratio between the SB of one component and the total SB for every galaxy. We take the mean of that ratio by only accounting for galaxies with non-zero SB. \textit{(c)}: The sample completeness of the mean ratio in (b). Lines indicate the fraction of galaxies which have non-zero SB at a given radius and thus contribute to the mean ratio in (b).}
        \label{fig:SB_mean}
    \end{figure*}
In this section we present our findings on the correlation between X-ray surface brightness and global intrinsic properties of our galaxies. We quantify the contribution of XRBs and SMBH sources to the SB as possible contaminants when determining properties of the CGM.

Throughout our investigation we split our full sample into the SF and QU sub-samples. In this way we probe different mechanisms responsible for maintaining a hot, X-ray-bright gas atmosphere in different galaxy populations.
Furthermore, we show SB profiles as a function of a normalized scale-free radius in order to make the sub-sample intrinsically more comparable irrespectively of differences in physical size. As a reference scale we choose the virial radius $\Rvir$.

We first investigate general properties of the QU and SF sample by constructing mean and median profiles of the full subsamples.
In figure \ref{fig:SB_mean}(a) we show the mean (thin lines) and median (thick lines) SB profiles of our complete sample accounting for every source component. The shaded area shows the 16-84 percentile ranges of the median profiles. The median profiles are always lower than the mean profiles because the latter is more sensitive towards extreme outliers as seen from the 84 percentile boundary.
Since we include every galaxy irrespective of stellar mass when constructing the mean and median here, we are naturally dominated by the brightest and presumably most massive galaxies. Additionally, the mean enhances the presence of substructures which is noticeable by the noisy behavior of the mean at large radii. The median is a more stable estimator here although it is less suited for capturing the SB in the outermost regions, where it is more sensitive to the large number of galaxies with zero emission.
The median profile drops to zero SB at $\sim 0.17\, \Rvir$ for the SF sample and at $\sim 0.22\, \Rvir$ for the QU sample. This means that less than 50\% of the galaxies have detectable SB beyond those radii, respectively. 
This can be visualized in figure \ref{fig:SB_mean}(c), where we show the sample completeness as the fraction of galaxies with non-zero SB at a given radius.
When comparing the two sub-samples, the SF sample is centrally brighter than the QU sample both in the median and mean. For radii larger than $0.1\,\Rvir$ the mean SB profiles of both samples are comparable with similar normalization and slope while median profiles are steeper for SF galaxies. For QU galaxies the median indicates a slightly more extended SB. According to figure \ref{fig:SB_mean}(c) this behavior is a result of more galaxies that are bright at larger radii.
We note that for large radii, the 16-84 percentiles of the QU and SF sample are identical which indicates that their CGM has similar properties.

In order to disentangle the contribution of different components to the SB profiles, we compute the ratio between the SB of each component and the total SB for each galaxy. 
This is shown in figure \ref{fig:SB_mean}(b).
In each radial bin, we then compute component-wise the mean of these ratios only accounting for galaxies with non-zero total SB.
In this way, we effectively reduce the number of available galaxies in our sample (cf. figure \ref{fig:SB_mean}(c)) especially at larger radii and can consequently inspect details of the SB profile in luminous galaxies. Different line styles in \ref{fig:SB_mean}(b) indicate the mean ratio of gas (dash-dotted), XRB (dashed) and SMBH (dotted) sources towards the total surface brightness of each galaxy, with SF and QU samples in the same colors as in panel (a).
We note that the mean ratio of each component is not biased by extreme outliers. We verified this by also computing the median ratio of each component, which yielded similar results.

Comparing these trends with the mean and median SB profiles from panel (a) of figure \ref{fig:SB_mean}, we conclude that the central increase of the SB in SF galaxies is mainly caused by an enhanced contribution from hot gas within $\lesssim0.05 \Rvir$. This is most likely due to the presence of a hot ISM where stellar feedback from active star-forming regions injects energy into the surroundings.
Conversely, the XRB component is more dominant for QU galaxies in the central regions. Due to the expected density distributions of stars and hot gas in quenched galaxies, the ISM contribution should be less pronounced in the center compared to XRB.
The contribution from SMBH sources is mostly insignificant except for the very center where every galaxy hosts a SMBH. We note, that the SMBH considered here would be X-ray faint due to our AGN exclusion.
Beyond $\sim0.05\,\Rvir$ both the SF and QU sample reach similar contribution levels in all components. Interestingly, the average XRB contribution is $\gtrsim10\%$ for radial bins close to $\Rvir$. We attribute this fact to the presence of stellar sub-structures.

Given the scatter in each sample, the mean SB profiles of the two classes of galaxies show little qualitative differences. The median SB however hints at SF galaxies being slightly less extended and having steeper profiles compared to QU galaxies.
Looking at relative contributions from different sources of the non-zero SB galaxies, we see that most of the difference is coming from the behavior of the ISM gas. In X-ray observations, the distinction between these two classes is also apparent on the ISM level \citep{Bogdan+2013, Kim+Fabbiano2015, Goulding+2016, Babyk+2018} and is supported by other independent simulation studies using the IllustrisTNG-100 (TNG100) suite \citep{Truong+2020}.

\subsection{Connection to galaxy properties}
In this section we will investigate the shape and slope of median SB profiles of our QU and SF galaxy subsample while accounting for differences in their global properties.
In the self-similar scenario, thermodynamic properties of the hot gaseous atmospheres are directly determined by the depth of the gravitational potential well of the underlying dark matter halo \citep[see e.g.][]{Sarazin1988}. Assuming that the main cooling mechanism is thermal bremsstrahlung, the X-ray luminosity $L_X$ of a halo can be expressed as
\begin{equation}
    L_X \propto \fgas^2 T_\mathrm{vir}^{0.5} M_\mathrm{vir}\, ,
    \label{eq:self-similar}
\end{equation}
with
\begin{equation}
    \fgas\lrb{<\Rvir} = \frac{M_{gas}\lrb{<\Rvir}}{M_\mathrm{vir}} \, ,
    \label{eq:fgas}
\end{equation}
as the gas fraction. Note, that the gas fraction is assumed to be constant with mass in the self-similar scenario, while the true gas fraction of a halo is strongly dependent on stellar and AGN feedback, as well as replenishing and depletion of the gas reservoir. These effects lead to deviations from the self-similar picture which are found in observations of elliptical galaxies as well as of star-forming galaxies \citep[][see review by]{Fabbiano2019}. 
Based on eq. \eqref{eq:self-similar}, we explore the gas fraction $\fgas$ and halo temperature $k_BT$ (estimated as in section \ref{sec:Determination}) of our galaxies and connect them to the SB profiles. 
We further inspect the relation to their total stellar mass $M_*$.

Measurements of the gas fraction in galaxies have been historically difficult and are typically limited to the innermost regions of the galaxy.
For instance, studies using survey data make use of stacking procedures to enhance the signal of weakly X-ray emitting gas in the outskirts which however makes quantitative statements on gas fractions in individual galaxies impossible.
Constraints on the hot gas fraction of individual galaxies are indeed sparse and mostly feasible for massive systems \citep[see e.g.][]{Bogdan+2013, Li+2017, Babyk+2018}.

In cosmological simulations, we are able to estimate directly the galaxies intrinsic gas fractions and connect them to X-ray properties.
    \begin{figure}
        \centering
        \includegraphics[width=\hsize]{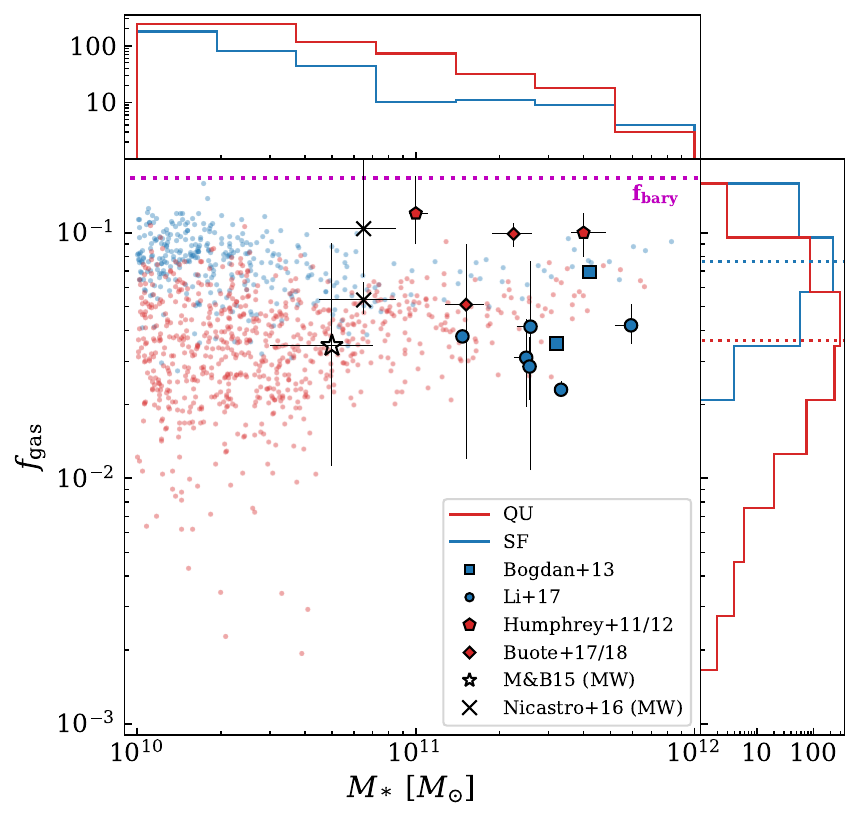}
        \caption{
            Gas fraction against stellar mass of our galaxy sample with $f_{gas}$ computed according to eq.\eqref{eq:fgas}. SF and QU galaxies are shown in blue and red respectively. Additional symbols with errorbars represent values obtained from the literature for comparison. The face-color of each symbol indicates SF / QU classification. The magenta line corresponds to the cosmic baryon fraction $f_{\rm{bary}} = 0.167$ adopted in the simulation.
        }
        \label{fig:fgas_mstar}
    \end{figure}   
In figure \ref{fig:fgas_mstar} we show the gas fraction for each galaxy in our sample against its stellar mass. Points are colored blue for SF and red for QU galaxies respectively. Additionally we show histograms of the stellar mass ($M_*$) and the gas fraction ($\fgas$) distributions of our sample which are attached to the respective axis. Dotted lines in the $\fgas$ histograms indicate the median value of the respective distribution. The magenta dotted line in the main panel shows the cosmic baryon fraction $f_{\rm{bary}}$ from the simulation.
For comparison, we also include measurements of gas fraction and stellar mass from galaxies in the local universe: NGC 720 \citep{Humphrey+2011} and NGC 1521 \citep{Humphrey+2012} (stars); NGC 1961 and NGC 6753 (squares) \citep{Bogdan+2013}; star-forming galaxies from \citep{Li+2017}; fossil group NGC 6482 \citep{Buote+2017} and compact elliptical galaxy Mrk 1216 \citep{Buote+2018}.
We selected these specific observational examples because the gas fractions were obtained from a detailed analysis of mass and density profiles resulting from deep X-ray observations. The mass profiles were then extrapolated to a radius of $R_{200c}$ to calculate the gas fraction.
Furthermore, we include gas fraction estimates of the MW \citep{Miller+Bregman2015, Nicastro+2016} which were obtained from modeling of \textsc{Ovii} and \textsc{Oviii} emission and \textsc{Ovii} absorption lines in the MW CGM, respectively and also quote the gas fraction at $R_{200}$. 
While our sample is consistent with the selected observations in terms of gas fractions, the observational values are biased towards X-ray bright galaxies which may not be representative of the average galaxy population.
In the simulated sample, QU galaxies have in general lower gas fractions than SF galaxies and span a larger range of values.
In particular, we note that the low-value tail of the gas fraction distribution in simulations is dominated by low-mass QU galaxies.
For comparison, we also report the cosmic $f_{\rm{bary}}$ value, which is as expected larger at all masses.

In figure \ref{fig:SB_prop} we report total SB profiles for the QU (upper panels) and SF (lower panels) samples, for different global intrinsic properties, i.e. $f_{\rm gas}$ (left),  $M_*$ (middle) and $k_B T$ (right).
Lines with different colors are median profiles binned by the respective property, using the same intervals for both QU and SF galaxies.
Given that the property bins do not contain equal numbers of galaxies, constraints would be less strong on the low-number bins. For comparison, we also report the median of the whole SF/QU subsample (dashed black line) from figure \ref{fig:SB_mean}.
For a better interpretation of observed profiles and for comparisons among different sizes, we define the quantity 
    \begin{equation}
    \label{eq:xi}
    \xi_i = \frac{300\,\rm{kpc}}{R_{\rm{vir},i}}\, .
    \end{equation}

We chose $300\,\rm{kpc}$ as a reference scale because it is close to the virial radius of a MW-mass halo and corresponds to the physical galacto-centric distance for which CGM emission was detected \citep{Comparat+2022, Chadayammuri+2022, Zhang+2024a}. 
This is used to highlight the distribution of sizes for the halos in each bin, visualized by the violins.
In Fig. \ref{fig:SB_prop}, we also report observed SB profiles for the BCG-like QU galaxies NGC 6482 and Mrk 1216 \citep{Buote+2017, Buote+2018} and for several local SF galaxies from \cite{Bogdan+2013, Bogdan+2015, Li+2017}. We color code the observational data points according to the considered property.

\subsubsection{Gas fraction}

We investigate the impact of the gas fraction on median SB profiles in the first column of figure \ref{fig:SB_prop}. 
Within the QU sample, median SB profiles have lower normalization and appear less extended with decreasing $\fgas$.
Their slope becomes slightly steeper with decreasing $\fgas$. For the lowest $\fgas$ bins, the median SB profiles drop to zero before $0.1\ \Rvir$. From the distribution of $\xi$ within each bin (violins), we can infer that more massive QU galaxies tend to have higher gas fractions. Compared to NGC 6482 and Mrk 1216, our sample has lower normalization.
In fact, NGC 6482 and Mrk 1216 are BCG-like and are expected to be brighter than normal elliptical galaxies \citep[see e.g.][]{Kim+Fabbiano2015}.

For SF median profiles, the normalization in the central regions decreases with decreasing $\fgas$. At the same time, galaxies with lower gas fractions seem to have more extended profiles. This is because the $\fgas$ bins are dominated by small galaxies which have less extended profiles. Visually, the SF median profiles seem to be steeper compared to the QU sample for the same $\fgas$ bins.
We caution however, that the interpretation of trends in the SF sample is more difficult here because of the narrower range in $\fgas$ compared to the QU sample. Since we use the same binning for both subsamples, this leads to fewer non-empty bins in the SF sample with similar size distributions $\xi$ among the bins.
A more quantitative approach regarding the steepness of the profiles w.r.t. the gas component will be shown in section \ref{sec:beta_prof} later on.
The median SB profiles of the SF sample are broadly consistent with estimates from observations of massive spiral galaxies \citep{Bogdan+2013, Li+2017}. As shown in Fig.~\ref{fig:fgas_mstar}, reported observations probe similar gas fractions compared to our SF sample but are more massive than the bulk of our SF galaxies. We also note that the annuli for which the SB was extracted in these observations are rather large and thus provide loose constraints.
The measured SB for massive spiral galaxies from \citet{Bogdan+2013} is higher than our median profiles which is due to the median being dominated by low mass SF galaxies.

\subsubsection{Stellar mass dependence}
\begin{figure*}
    \centering
    \includegraphics[width=\hsize]{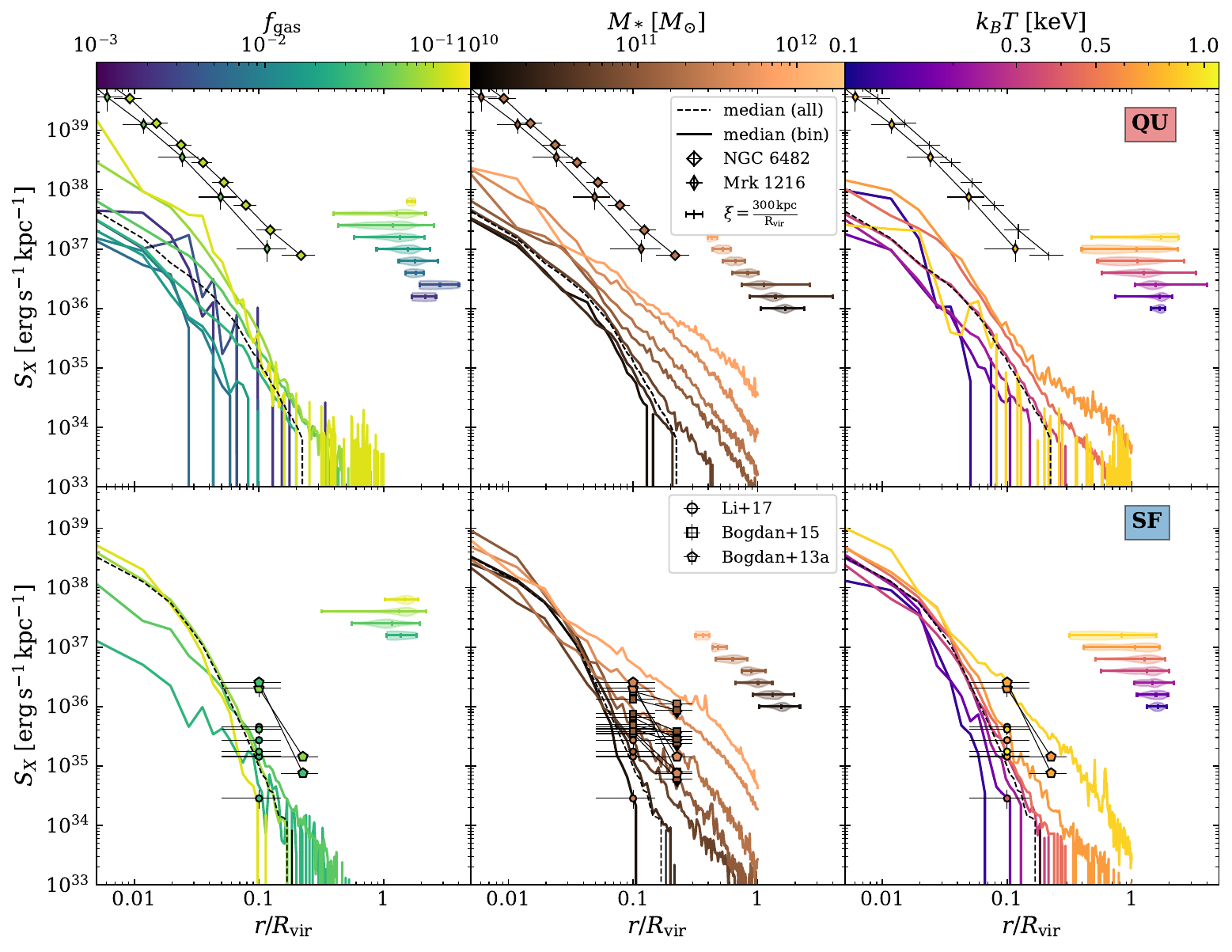}
    \caption{
        Scale-free median SB profiles of quiescent (\textit{top}) and star-forming (\textit{bottom}) galaxies. Galaxies are binned by gas fraction $\fgas$ (\textit{left}), stellar mass $M_*$(\textit{center}), and emissivity weighted temperature $k_BT$ (\textit{right}). Colors indicate the central value of each bin for the respective quantity. We include SB profiles for NGC 6482 \citep{Buote+2017} and Mrk 1216 \citep{Buote+2018} in the QU panels and measurements of the extended emission in SF galaxies from \citep{Bogdan+2013, Bogdan+2015, Li+2017} for the SF panels. The violin plots indicate the distribution of $\xi$ (eq. \eqref{eq:xi} within each quantity bin. The horizontal extent of each violin indicate the minimum and maximum value of $\xi$ within the respective bin. The central tick indicates the mean value of $\xi$. The height of each violin is proportional to the number density of $\xi$ in the bin. The black dashed line is the median profile of the QU and SF sub sample respectively.
        }
    \label{fig:SB_prop}
\end{figure*}

In the central column of figure \ref{fig:SB_prop}, we show the stellar mass dependency of median SB profiles.
The normalization of median profiles in the QU subsample increases with increasing stellar mass. The slope of the profiles remains mostly unaffected by changes in stellar mass and the extent of the profiles decreases with decreasing stellar mass. The distribution in $\xi$ confirms that larger galaxies have higher stellar mass. The median profiles of all stellar mass bins in the QU case are below the observational sample for the same reason as before.
For median SF profiles the normalization appears to be unaffected by stellar mass. Instead, decreasing stellar mass leads to a steepening of the profiles. The distribution in $\xi$ for each bin shows the same trend as in the QU case. Our median SF profiles are in agreement with observational results in the $0.05-0.15\ \Rvir$ radial range. For the $0.15-0.3\ \Rvir$ radial range, our data is consistent with observations. However, the high stellar mass galaxies from \citet{Bogdan+2013} are below our median profiles for a similar stellar mass bin while the slightly less massive galaxies from \citet{Bogdan+2015} are above median profiles for similar stellar mass bins in the $0.15-0.3\,\Rvir$ radial range. This is likely caused by a selection effect of the observational sample compared to our statistical sample.
The stellar mass dependency of galaxy X-ray luminosity has been well studied in recent years. Clear correlations have been found for elliptical and spheroidal galaxies with the integrated stellar light \citep{Kim+Fabbiano2013}, for elliptical galaxies with their dynamical mass \citep{Kim+Fabbiano2015, Forbes+2017}. For star-forming galaxies, \citet{Aird+2017} showed a connection between the stellar mass and mode of X-ray luminosity.
Recent results from the EFEDs field of eROSITA also indicate the presence of a correlation between stellar mass and the SB normalization \citep{Comparat+2022, Chadayammuri+2022, Zhang+2024a}.

Generally, a stellar mass trend in the SB should be expected as it can be connected to the gas mass through halo mass relations. The steepening of the profiles in the SF sub sample indicates a change in the gas density distribution depending on total stellar mass, since the profiles are dominated by gas emission in the outskirts (see fig. \ref{fig:SB_mean}). We will quantify this effect and highlight differences between the QU and SF subsamples in section \ref{sec:beta_prof}.
    
\subsubsection{Temperature}
In the right column of figure \ref{fig:SB_prop} we show the temperature dependence of median SB profiles.
With increasing temperature, the median profile of the QU sample in the top panel shows an increase in normalization and becomes more extended. 
Except for the highest temperature bin, the $\xi$ distribution indicates that larger galaxies have higher temperatures. The median profile of the highest bin lies in between the other bins in terms of normalization and extent and mostly consists of galaxies with large $\xi$, and thus of smaller halos. This is not expected from empirical scaling relations, where higher temperatures are associated with more massive galaxies \citep[see e.g.]{Kim+Fabbiano2015, Goulding+2016, Babyk+2018}.
Upon inspection, 4 out of the 6 galaxies within the highest temperature bin showed no gas emission and some XRB emission in the central 0.1 $\Rvir$. They have low stellar mass ($<2\cdot10^{10}\,\Msun$) and slightly lower $\fgas$ ($\eqsim 0.06$).
Outside of 0.1 $\Rvir$, they have very shallow gas SB profiles with low normalization and little XRB contribution. We thus argue, that some recent event removed the central gas of these galaxies and simultaneously heated their gaseous atmosphere at larger radii.
The SF sample shows a weak increase in normalization and radial extent with increasing halo temperature. With decreasing temperature the profiles again become steeper because the low temperature bins consist of more small galaxies.
The distribution in $\xi$ consistently indicates higher temperatures for larger halos. The SF sample is again in agreement with observations in terms of normalization for all the radial ranges shown. In particular, one galaxy in the sample from \citet{Li+2017} has a higher temperature than the highest temperature bin. Again we argue that differences between properties are a result of selection effects in observations.

\subsection{SB profiles of the gas component}
As shown in Fig.~\ref{fig:SB_mean}, the hot gas is responsible for most of the X-ray emission in our AGN-cleaned sample of galaxies, throughout the majority of the galaxy volume.
It is thus interesting to further inspect the gas SB properties separately.
This is directly possible in simulations, where we can predict for each galaxy the emission of the contaminating components (i.e. SMBH and XRB) individually.
In observations, the study of the hot gas distribution of galaxies is more difficult, due to uncertainties regarding contaminants such as the central emission of the SMBH, point sources in the galactic field, background modeling and instrumental response.
Nonetheless, studies of the CGM emission can be attempted through stacking of galaxy spectra, as shown by \citep{Oppenheimer+2020}, based on mock observations of simulated galaxies extracted from IllustrisTNG and EAGLE simulations.
In fact, recent observational studies based on eROSITA data, successfully employed stacking to find emission above the background level from the CGM of MW/M31-like galaxies \citep{Comparat+2022, Chadayammuri+2022, Zhang+2024a}.

\begin{figure}
    \centering
    \includegraphics[width=\hsize]{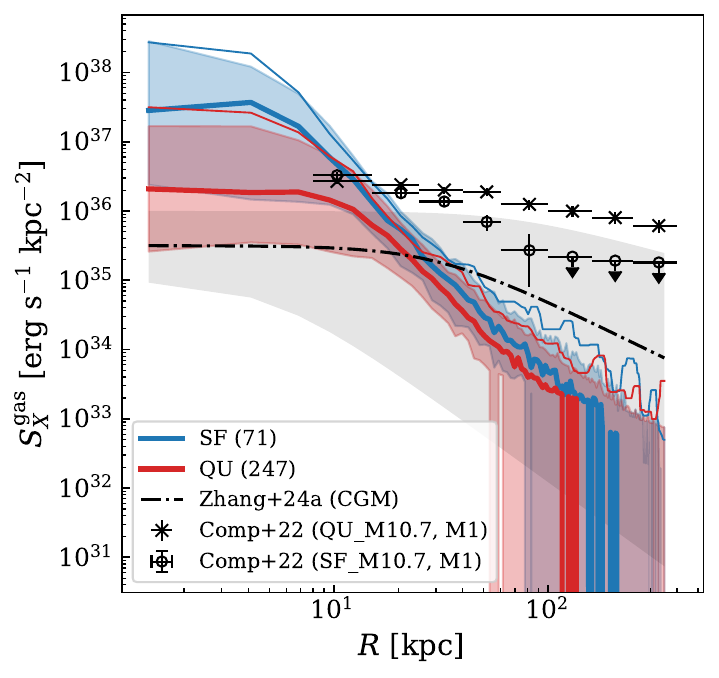}
    \caption{X-ray SB profile of the hot gas component recreating the mass cuts from \citet{Comparat+2022}. We recreate the observational sample from eFEDs galaxies \citep{Comparat+2022} by replicating their M1 mask for the SF\_M10.7 and QU\_M10.7 mass bins and show their background subtracted results. Thin and thick solid colored lines are mean and median SB profiles of our sample of galaxies. We apply mass cuts on our sample ranging from $M_*=10^{10.46-11}\,\Msun$ for QU galaxies and $M_*=10^{10.48-11}\,\Msun$ for SF galaxies and report the total number of galaxies in each stack in the legend. We apply a median filter on the mean stacked profiles (thin solid) to remove substructures. Additionally we show the best-fit SB profile for the CGM of MW-mass galaxies from \citet{Zhang+2024a} which probes a similar stellar mass range.}
    \label{fig:Compare}
\end{figure}

Here, we confront our findings with these recent eROSITA results,as shown in figure \ref{fig:Compare}. In particular, we report the stacking results for the QU\_M10.7 and the SF\_M10.7 sample from the eFEDs field \citep{Comparat+2022} (Comp22).
We use the same mass selection as their M1 mask which removes the signal from bright AGN and is comparable to our AGN cleaned sample. The QU\_10.7 sample consists of 7267 quiescent ($\log(\rm{sSFR\,[\rm{yr^{-1}}]}) < -11$) GAMA matched galaxies in the mass range of $M_*=10^{10-11}\,\Msun$ and the SF\_M10.7 sample with 9846 star-forming galaxies in the mass range $M_*=10^{10.4-11}\,\Msun$. The mass ranges were chosen such that the mean stellar mass in each sample is $M_*=10^{10.7}\,\Msun$. The average redshift is 0.2 and 0.23 for QU\_M10.7 and SF\_M10.7 respectively. 
In figure \ref{fig:Compare}, we also include the best fit $\beta$ model for the CGM of 30825 eROSITA stacked central galaxies in the MW mass range ($M_*\sim10^{10.5-11}$) and a median redshift $z\sim0.08$ from \cite{Zhang+2024a} (Z24a) (dash-dotted black line). The associated shaded area corresponds to the $1\sigma$ uncertainty in the best-fit parameters of their $\beta$-model. The contamination of XRBs, AGNs and satellite galaxies was accounted for through empirical modeling and the fitting was performed on the background-subtracted stacked SB profile. 
Furthermore, they do not distinguish between star-forming and quiescent galaxies such that both are included in their mass range. An improvement w.r.t Comp22 is a more detailed treatment of contamination from satellite galaxies which leads to a steeper profile.
We note that their best-fit $\beta$-model for the MW mass regime has large errorbars in general but is representative of their full gas profile.

For a more faithful comparison, we restrict here to a subsample of our simulated galaxies that more closely resembles the observational selection.
By applying exactly the same mass ranges as the QU\_M10.7 and SF\_M10.7 on our galaxy sample, we obtain 93 SF and 645 QU galaxies, which does not reflect the same galaxy number ratio and fails to give the correct mean mass of $\bar{M}_*=10^{10.7}\,\Msun$.
We thus define our mass ranges such that $\bar{M}_*=10^{10.7}\,\Msun$, which results in 71 SF ($M_*=10^{10.48-11}\Msun$) and 247 QU ($M_*=10^{10.46-11}\Msun$) galaxies.
We did not try to recreate an exact match of their stellar mass distribution (see Table 1 in Comp22), because we can not account for the redshift distribution given the fixed redshift of our simulation box.
Nonetheless, our chosen mass ranges do in fact overlap with the mass ranges and mean redshift presented in Z24a. 

In figure \ref{fig:Compare}, we plot the mean (thin solid) and median (thick solid) SB profiles of the gas component for our $\bar{M}_*$-matched sub-sample, in blue for SF galaxies and red for QU galaxies. The shaded area corresponds to the 25-75 percentiles and we additionally apply a median filter on the mean profiles to mask out satellite contribution.
Compared to the profiles by Comp22, we find overall steeper profiles for both QU and SF galaxies.
We note nonetheless that this difference is stronger for the QU subsample, as the SF mean profile by Comp22 indicates a moderatly steeper trend in the central bins (up to $\sim 100 kpc$) but is mostly constrained by upper limits in the outer regions.
While hinting at a flatter shape than simulations, no strong conclusions on the comparison can be therefore drawn for the SF subsample. Furthermore, the procedure in Z24a showed that satellite contamination in Comp22 may be significant.
We find that our mean and median SB profiles are in good agreement with results from Z24a, given the uncertainties.
In the central 10 kpc, our median SB profiles appear generally brighter than in Z24a. Nevertheless, the QU median stack is compatible with results from Z24a within the 25 percentile.
Another detail in the observational analysis is the treatment of point sources and nuclear emission in each galaxy.
Typically, excess in nuclear emission is attributed to SMBH activity and is consequently removed from the analysis. This could lead to an underestimation of the central SB in observations if the emission is originating from the gas component instead of an AGN. 

\subsubsection{Beta profiles of the gaseous component in individual galaxies}
\label{sec:beta_prof}
\begin{figure}
        \centering
        \includegraphics[width=\hsize]{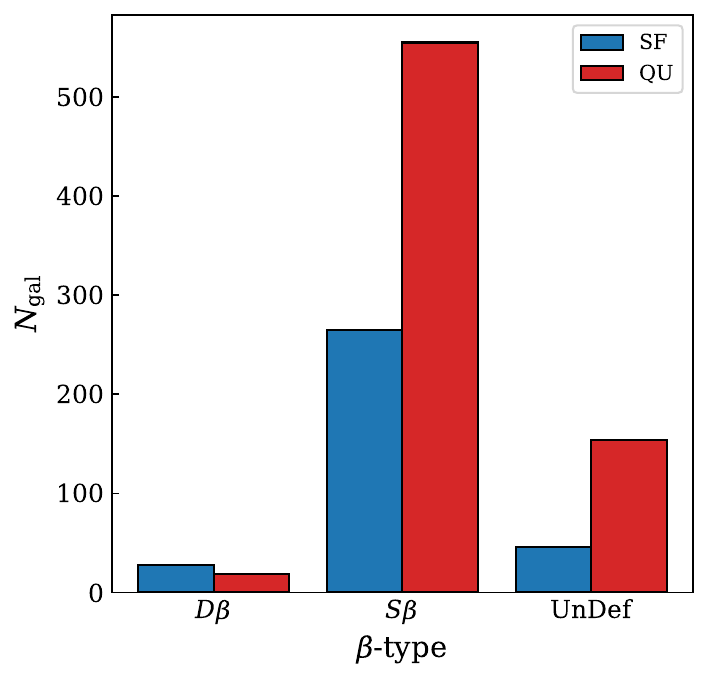}
        \caption{Result of the labeling process after fitting each SB profile with a single and double $\beta$ profile. The main classification criterion is based on the reduced $\chi^2$ as well as parameter degeneracy (see text).}
        \label{fig:beta_label}
    \end{figure}

    The shape of the SB profile associated to the gas component can be further inspected by modeling it with a $\beta$ profile, as often done in observations.
    Specifically, we model the SB profiles of the hot gaseous component by fitting both a single $\beta$-profile ($S\beta$) with 3 free parameters, 
    \begin{equation}
        \label{eq:beta_prof}
        S_X(r) = S_0\lrb{1+\lrb{\dfrac{r}{r_c}}^2}^{0.5-3\beta}\, ,
    \end{equation}
    where $S_0$ is the normalization at $r=0$, $r_c$ is the core radius and the slope $\beta$ \citep{BetaProf1978}, and a double $\beta$-profile ($D\beta$) with 6 free parameters. The $\beta$-profile assumes spherical symmetry of the gas density distribution where the gas is in isothermal equilibrium within the gravitational potential of the galaxy.
    We use a standard $\chi^2$ least-square algorithm in $\log$-space with equal-size radial bins in units of $\Rvir$ for each halo to fit the SB radial profile. We assume Poissonian uncertainties on the SB in each radial bin based on the photons counts.
    In order to account for substructures in the halo outskirts during the fitting procedure, we apply a median filter on the SB profiles which removes sharp spikes in SB and effectively functions as a mask. 
    We use the reduced $\chi^2$ value to determine the best fit.
    If the reduced $\chi^2$ of single and double $\beta$-profile fits are both close to 1, we prefer the single $\beta$-profile, which has less free parameters.
    If the best fit is a double $\beta$-profile but relative uncertainties in the fitting parameters are large $\lrb{\frac{\Delta x}{x}>0.8}$ due to degeneracies, and the single $\beta$-model also yields a good fit, we prefer the latter.
    In cases where neither a single nor a double $\beta$-profile adequately describe the data, we label the galaxy as an "undefined" case, and do not consider those for the subsequent analysis. In figure \ref{fig:beta_label} we show the results of this fitting process.
    The blue and red colored histograms show the distribution of QU and SF galaxies.
    Most galaxies in our sample are consistent with a $S\beta$ profile. 
    The most massive galaxies in our sample are instead better described by the $D\beta$ model.
    We show exemplary profiles of both categories in appendix~\ref{app:examples} (\ref{fig:beta_example}).
    The undefined cases are galaxies which do not have any surface brightness or have too few (<10) non-zero radial bins, and are exclusively found at low halo masses. 
    
\begin{figure*}
        \centering
        \includegraphics[width=\hsize]{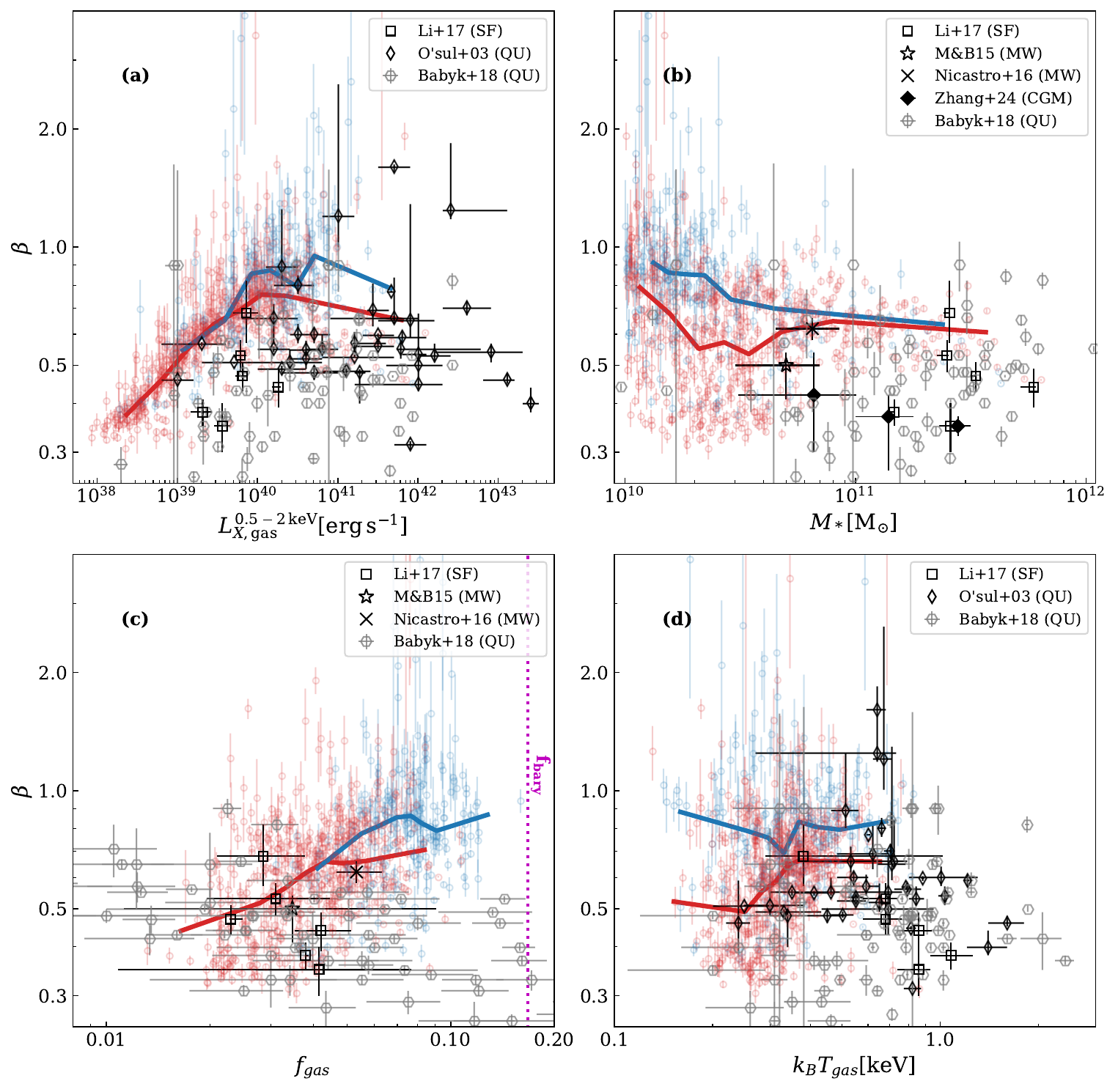}
        \caption{Best-fit slope $\beta$ of each galaxy's gas SB profile labeled as a $S\beta$ profile (eq. \eqref{eq:beta_prof}) against various halo properties within $\Rvir$: \textit{(a)} gas luminosity $L_{X, gas}$; \textit{(b)} stellar mass $M_*$ obtained from stellar resolution elements bound to the parent halo; \textit{(c)} gas fraction $f_{\rm{gas}}$ derived from gas resolution elements bound to the parent halo excluding star-forming and low-temperature ($<10^5\,\rm{K}$) gas. In panel \textit{(c)} the \textit{dotted magenta} line indicates the cosmic baryon fraction in the simulation; \textit{(d)} emissivity weighted hot gas temperature $T_{\rm{gas}}$. The exact retrieval of these quantities are outlined in section \ref{sec:data_set}. The thick solid line in each panel indicates the median value of $\beta$. For comparison we include the sample of massive elliptical galaxies from \citet{Osul+2003} (O'sul+03) and massive star-forming galaxies of \citet{Li+2017}. Additionally, we compare to $\beta$ models of the MW from \citet{Miller+Bregman2015} (M\&B15) and \citet{Nicastro+2016} (model A) in (b) and (c). 
        }
        \label{fig:beta_corr}
    \end{figure*}

    As a second step, we investigate the relation between the shape of the density profile (quantified via the slope $\beta$ of the $\beta$-profile) and global properties of the galaxies. To this scope, we restrict our analysis to the subsample of galaxies that are best modeled by a $S\beta$ profile.
    In figure \ref{fig:beta_corr}, we show the best-fit single slope $\beta$ as a function of total gas luminosity $L_{X, \rm{gas}}$, stellar mass $M_*$, hot gas fraction $\fgas$ and emissivity-weighted hot gas temperature $k_B T$, in panels (a), (b), (c) and (d) respectively. 
    The simulation data points, with error bars, are marked in red and blue to distinguish QU from SF galaxies respectively.
    The thick colored lines represent the median $\beta$ of the respective sample for equal-count bins.
    
    In general, we find that the SB profiles in the SF sample have steeper slopes compared to the QU sample in all examined properties. Furthermore, uncertainties on the slope increase for larger values of $\beta$, due to degeneracy with the core radius $r_c$. We also note that the overall scatter is large and the two subsamples have significant overlap.
    
    In panel (a) we find a strong positive correlation between the slope of the SB profile and $L_{X, \rm{gas}}$ up to $L_{X, \rm{gas}} \approx 5\cdot10^{40}\,\ulum$ for QU galaxies and $L_{X, \rm{gas}} \approx 10^{41}\,\ulum$ for SF galaxies. This indicates that the gas emission becomes more centrally concentrated for more luminous halos in these cases.
    At higher luminosities, the median slope levels between $\beta \sim 0.6$ (for QU galaxies) and $0.8$ (for SF galaxies).
    Panel (b) indicates that the simulated galaxies span the largest range in $\beta$ values at the lowest stellar masses. Despite the large scatter, in this stellar mass regime we still find a moderate tendency for steeper profiles in SF galaxies.
    For stellar mass above $10^{11} \Msun$, $\beta$ remains around $\sim 0.6$, for both the SF and QU sample.
    In panel (c) we find that $\beta$ is positively correlated with $\fgas$ for both galaxy samples. We also notice that the median slopes of the two subsamples seem to connect across the full $\fgas$ range.
    This suggests that the distribution of hot gas tends to be more centrally-concentrated in galaxies with higher gas fractions.
    A small fraction of the central increase can be attributed to a hot ISM component in SF galaxies due to SN feedback from newly formed stars. However, in panel (d) we do not find a clear correlation for steeper slopes and the temperature of the halo.
    Since the slope is most sensitive to SB outside the core region, this suggests that the steep profiles are an intrinsic property of the halo.

    In order to better interpret the origin of the largest $\beta$ values, we inspect directly the corresponding profiles. We find that these galaxies either have also large uncertainties on the core radius $r_c$, or present sharp drops in SB at $r\gtrsim 0.1\Rvir$.
    The latter corresponds to a few effective radii for those galaxies.
    Since we found the steepest slopes in low-mass galaxies, the compactness of the profiles may be caused by resolution limits in the simulation. In this case the low density of the halo outskirts is represented by a few resolution elements only which leads to stochastic effects.
    At the same time, the smallest values of $\beta$ are also found in low-mass systems, and this is likely a result of extreme feedback events displacing the gas beyond the galaxy boundary.

    In order to cross-validate our statistical results with observations, we compare to several observational studies. In all panels we show $\beta$ for a sample of nearby QU galaxies (gray hexagons), where X-ray properties were obtained from deep CHANDRA observations including extensive modeling of background sources and contaminants \citep{Babyk+2018}.
    Their studied sample also includes BCG and cG galaxies which we exclude for the comparison. Luminosities and temperatures were directly taken from the aforementioned study. We inferred stellar masses from the central stellar velocity dispersion given for each galaxy in their study using scaling relations from \citet{Zahid+2016}. Gas fractions were obtained from the given gas masses and the given dynamical mass.
    We note, that X-ray quantities derived in \citet{Babyk+2018} were extracted within five effective radii which is smaller than the $\Rvir$ in our analysis. While the given profile slopes and halo temperatures should not change significantly at larger radii, properties such as stellar mass and gas fraction are likely lower bounds.
    In panels (a) and (d) we additionally show a different sample of massive elliptical galaxies (black empty diamonds), by \cite{Osul+2003} who used ROSAT data to obtain SB profiles of massive elliptical galaxies.
    In panel (b) we include best fit slopes for the CGM of MW-mass and M31-mass galaxies from \citet{Zhang+2024a} (black filled diamonds) derived from stacking analysis of the first eROSITA full-sky survey (eRASS:1). We also show a sample of massive SF galaxies in all panels from \citet{Li+2017} who used XMM-Newton data for their analysis.
    In panel (b) and (c) there are also values for the Milky Way (MW) derived by \cite{Miller+Bregman2015} and \cite{Nicastro+2016}. 
    The data by \cite{Miller+Bregman2015} result from a symmetric $\beta$-like profile which has been flattened along the axis perpendicular to the galactic disk, and uses XMM-Newton data of $\textsc{Ovii}$ and $\textsc{Oviii}$ emission lines from the MW CGM. 
    The values from \cite{Nicastro+2016} (model A in their study) refer to a true spherical symmetric profile derived from X-ray absorption lines in Chandra data associated to the MW CGM.
    We note that our slopes for the SF sample are systematically larger than the slopes from \citet{Li+2017} and compared to the MW. 
    Additionally, the gas fractions of the MW and the sample from \citet{Li+2017} are lower than those of our SF sample. Interestingly, the slopes from \citet{Li+2017} seem to agree better with our QU sample, especially in terms of $L_{X,gas}$ and $\fgas$ but are still on the lower side. 
    Our derived slopes are in broad agreement with the $\beta$ values derived in \citet{Babyk+2018}, despite being typically higher.
    While luminosities in their study are similar to ours, they have more high mass galaxies. Especially in the high stellar mass regime, their SB profiles are shallower. Regarding gas fractions, their sample has a larger range compared to ours. Interestingly, their sample shows a decline in slope with increasing $\fgas$ which is in contrast to our results.
    Furthermore, they probe higher temperatures compared to our analysis which is connected to the higher stellar mass of their galaxies.
    In general, the temperature of the gaseous halo in observed galaxies is consistent with our sample and also does not show a clear correlation with $\beta$.
    The sample from \citet{Osul+2003} is consistent with slopes derived for our galaxy sample and also contains a few galaxies with $\beta \gtrsim 1$.
    In their case, larger slopes are likely connected to the environment of their galaxy sample. Most of their targets lie in a cluster or group environment which can in principle affect X-ray properties of the galaxy even after accounting for the cluster emission in spectral modeling.    

\subsection{Global X-ray luminosity}

\begin{figure*}
    \centering
    \includegraphics[width=\hsize]{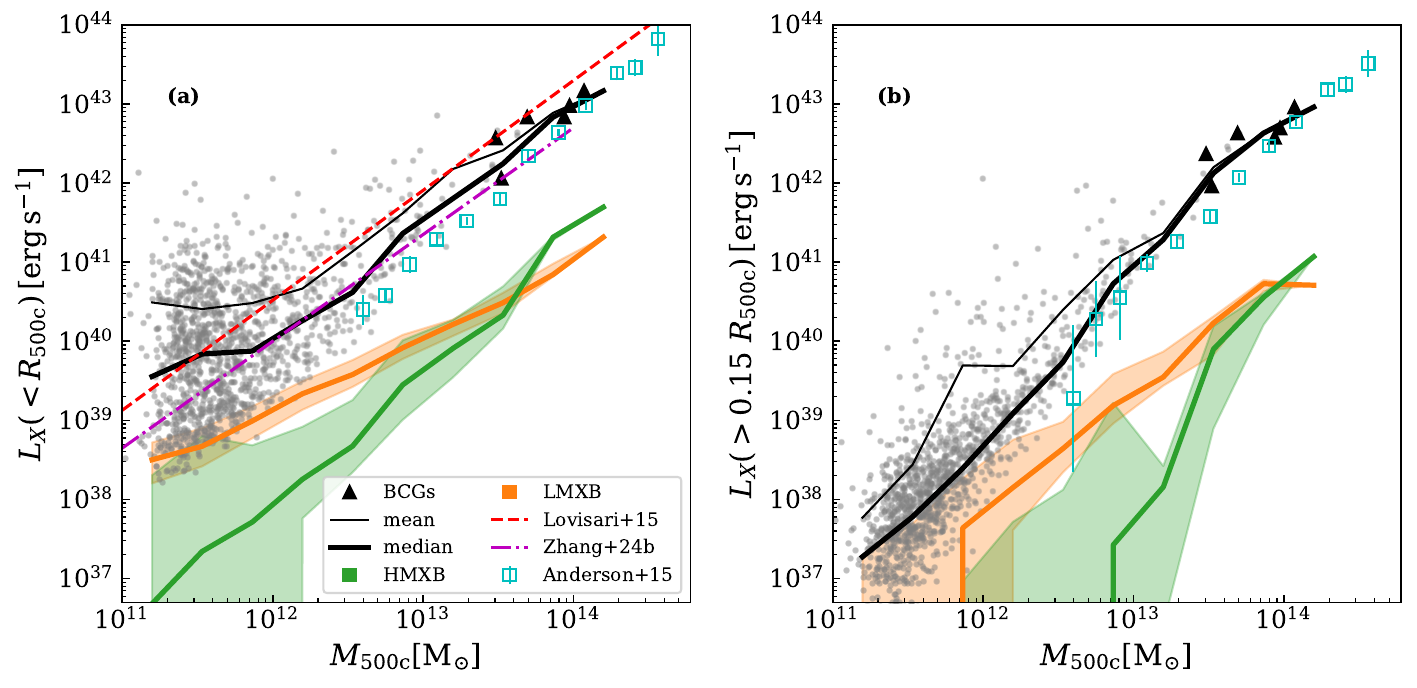}
    \caption{Total X-ray luminosity as a function of halo mass ($M_{500c}$)
    \textit{(a)} within $R_{500c}$ of each galaxy, \textit{(b)} within $(0.15-1)\,R_{500c}$. 
    Grey dots represents all galaxies in our full sample, including the AGN systems, with BCG galaxies marked as black triangles.
    Thin and thick lines represent the mean and median luminosity of our sample, respectively. The contribution from HMXBs and LMXBs in our sample is shown in \textit{orange} and \textit{green} together with the 16-84 percentile range as the shaded area. Additionally, we show data from \citet{Anderson+2015} (\textit{cyan squares}) for the total X-ray luminosity within $R_{500}$ in \textit{(a)} and CGM luminosity within $0.15-1\,R_{500}$ in \textit{(b)}. The sample consists of central galaxies and results from a stacking analysis using ROSAT data. Their total mass is derived from forward-modeling of the $L_X-M_{500c}$ relation of gas dominated halos in their sample. The magenta dash-dotted line is the best fit $L_X-M_{500c}$ relation for stacked galaxies in eRASS:4 from \citet{Zhang+2024b}. The red dashed line shows the bias-corrected best fit $L_X-M_{500c}$ relation from \cite{Lovisari+2015}.}
    \label{fig:lx_mstar}
\end{figure*}

In figure \ref{fig:lx_mstar} we show total X-ray luminosities of our complete sample (including AGN) as a function of the total mass $M_{500c}$ (gray dots) and compare to scaling relations from \citet{Anderson+2015} (cyan boxes),  \citet{Lovisari+2015} (red dashed line) and \citet{Zhang+2024b} (magenta dash-dotted line).
We additionally include 6 group-like halos with $M_*>10^{12}\,\Msun$ ($M_{\rm{500c}}>5\cdot10^{13}\,\Msun$) from the same simulated volume in Fig \ref{fig:lx_mstar} and highlight them as black triangles.
For the total luminosity of our sample, we combine the emission of hot gas, XRBs and SMBHs. 
Instead of considering the whole $R_{\rm vir}$ extent for the simulated galaxies, we extract the properties in the same regions used by \citet{Anderson+2015}, namely within $R_{500c}$ and $[0.15$--$1]\,R_{500c}$.
They obtained their SB measurements from a bootstrapped stacking procedure with data from the Rosat All-Sky Survey (RASS) of SDSS (Sloan Digital Sky Survey) confirmed galaxies in a stellar mass range of $10^{10-12}\,\Msun$.
Luminosities $L_{\rm{tot}}$ and $L_{\rm{CGM}}$ in their study were extracted from stacked SB profiles of central galaxies. They derive total masses for their stellar mass bins by forward-modeling of the $L_X-M_{500c}$ relation using $L_{\rm{tot}}$ from their stacks. With this approach, they did not attempt to derive total masses for halos with $M_*<10^{10.8}\,\Msun$ ($M_{\rm{500c}}<10^{12.4}\,\Msun$) due to significant contamination from XRBs.
We note, that \citet{Anderson+2015} referred to the radial range $[0.15$--$1]\,R_{500c}$ as CGM, which we also adopt here for convenience.
The best fit $L_X-M_{500c}$ relation from \citet{Lovisari+2015} is accounting for selection bias and was derived from a sample of galaxy groups and clusters using \textit{XMM-Newton} observations.
The relation from \citet{Zhang+2024b} results from a stacking analysis of central galaxies in eRASS:4 and accounts for source contamination from a central SMBH and XRBs.
Thick solid lines in figure \ref{fig:lx_mstar} represent the median of our sample. Colored lines and shaded area represent the contribution of HMXBs (green) and LMXBs (orange) together with their 16-84 percentile which is a direct prediction of our XRB model \citep[see][for details]{SVZ:I}. We are thus also able to provide constraints of XRB contribution for the CGM regime.
The thin black line is the mean of our sample.

Generally, the median total luminosity of our galaxy sample within $R_{500}$ (panel (a)) is in very good agreement with the reported scaling relations from the literature. 
At intermediate masses, $2 \cdot 10^{12}\lesssim M_{500c}\, [\Msun]\lesssim 10^{13}$, the simulated median relation naturally shows increasing deviations from the relation by \citet{Lovisari+2015}, whose sample does not include low-mass systems.
At lower masses, $M_{500c}\lesssim 2\cdot 10^{12}\,\Msun$, simulation data show a large scatter in luminosity but we still find a broad agreement with the observed relations.
The increasing scatter flattens the mean and median at these low halo masses, and is driven by extreme outliers especially at high luminosities.
We note, that the scaling relations from \citet{Zhang+2024b} and \citet{Lovisari+2015} have very similar slopes, despite being derived for vastly different halo masses. This hints at a common mechanism shaping the overall matter distribution of all halos.

The median LMXB luminosity follows a almost linear relation with $M_{500c}$. By construction, LMXB contribution should linearly increase with stellar mass of the galaxy \citep{SVZ:I}. Deviations from a linear relation with $M_{500c}$ arise from a non-linear stellar mass function. For the lowest luminosities in our sample, the emission from LMXBs dominates with respect to HMXBs, although the major emitting component remains the hot gas.
The median HMXB luminosity increases with halo mass, but its contribution to the total $L_{X}$ is significantly lower compared to LMXBs, except for the highest halo masses.

We note however, that HMXB contribution is highly dependent on the sample and model assumptions.
In A15 for instance, they use a SFR indicator which probes the star formation of the past 300 Myr while we probe star formation in the past 100 Myr. Furthermore, they employed simple scaling relations between SFR and HMXB luminosity while we directly sample the HMXB luminosity functions \citep[see][]{SVZ:I}.
The latter introduces Poissonian noise in the $L_{\rm{HMXB}}$ scaling relation which steepens the slope of the relation at low SFR \citep{Gilfanov+2004b, SVZ:I}.
Moreover, A15 argue that the population of star-forming galaxies increases with lower stellar mass. While this is true in our case as well (see figure \ref{fig:fgas_mstar}), each stellar mass bin has nonetheless more quiescent galaxies than star-forming galaxies.
We also inspected the main-sequence for our galaxy sample (shown in figure \ref{fig:mainseq}, in appendix~\ref{app:mainseq}) and found that there are indeed few SF galaxies which are close to the main-sequence at low stellar mass.
Hence, we expect SF galaxies below the main sequence to have less contribution from HMXBs, given their SFR, compared to a linear relation.

The mean and median CGM luminosity of our sample (panel (b)) is in excellent agreement with stacking results of A15 (cyan) where we used the same values for $M_{500c}$ as in panel (a).
The contribution of LMXBs and HMXBs to the total CGM luminosity shows steeper trends compared to the total luminosity within $R_{500}$. We argue that there is more substructure in the outskirts of more massive halos which amplifies the contribution from XRBs.
Moreover, especially LMXBs can contribute significantly towards the CGM luminosity of the least luminous galaxies at low halo masses. In these cases the hot gas fraction is close to zero and residual LMXB emission must either be associated to satellites or a diffuse stellar component outside the central $0.15 R_{500c}$. 
We note, that the median and mean CGM luminosity for low mass halos ($M_{500c}<10^{12}\,\Msun$) is significantly lower than the total luminosity by at least an order of magnitude. For larger masses, the CGM luminosity becomes comparable to the total luminosity.
This implies that most of the emission is centrally concentrated in our sample, which is in line with steeper $\beta$-profiles from section \ref{sec:beta_prof}.

\section{Discussion}
\label{sec:Discussion}
    \subsection{Inclusion of absorption}
    By including a weak foreground absorption to model the idealized emission assigned to each gas element, we effectively reduce the X-ray emission in the soft part.
    Therefore the luminosity retrieved directly from the photon counts, without modeling the spectrum, is in principle a lower limit to the true intrinsic luminosity, even if no instrumental response is included and the ideal photon emission is employed. 
    However, we verified that
    the foreground absorber does not significantly impact the derived luminosity for each halo, by conducting a simple experiment.
    We again applied unit 1 of the \phox{} algorithm on a sub-volume of the simulation box without the weak foreground absorption component. All other settings were left the same as described in section \ref{sec:data_set}.
    We compare the total number of photons produced in the case with no absorption and the original setup for the same sub-volume in the energy range 0.5-2 keV.
    Since we take into account all particles within the simulated sub-volume, we probe a wide range of temperatures and metallicities being affected by absorption.
    Ultimately, the foreground absorption component, with column density $N_H=10^{20}\,\rm{cm^{-2}}$, reduces the total number of photons by $\approx5\%$ in the considered sub-volume.
    We thus recompute the luminosities as in section \ref{sec:Results} and find that the true intrinsic luminosity is typically underestimated by up to $5\%$ when absorption is included.

    Another point worth considering is the inclusion of an intrinsic component mimicking ISM absorption in SF galaxies which we did not include in this study. Typical values for ISM absorption in observed SF galaxies are of the order of $\sim 10^{21}\, \rm{cm}^{-2}$ \citep[see e.g.][]{Lehmer+2022}, which would lead to a reduction of SB in the central 0.1 $\Rvir$ in the gas component of SF galaxies by $\gtrsim 30\%$, ultimately reducing the normalization of the profiles. 

    \subsection{Beta profiles}
    We chose a least-squares algorithm in log-space to fit the SB profiles with a single or double $\beta$-model. We used the reduced $\chi^2$ as the main criterion
    for comparison, to decide between a $S\beta$ and $D\beta$ model. We manually validated the resulting choice for all galaxies in the AGN cleaned sample and confirmed if the fits reached convergence.
    Additionally, we cross-validated the resulting best-fit parameters using other fitting techniques.
    On the most massive galaxies within our sample ($M_*>5\cdot10^{11}\, \Msun$), we employed a likelihood minimization using a C-statistic \citep{Cash1979}, a gradient-descend algorithm as implemented in \texttt{scipy} \citep{2020SciPy-NMeth} and a Markov-Chain Monte-Carlo (MCMC) approach with flat and informed priors using the software package \textsc{emcee} \citep{emcee}.
    Both the C-statisic and the gradient-descend method yielded similar best-fit values and agreed with the least-squared method. The MCMC approach was sensitive to the assumed priors and slow to converge. It yielded median fit-parameters that were more degenerate than the ones found with the other methods and were systematically lower by $\sim 10\%$. 
    Given these results, we decided to rely on the least-square results.
    
\begin{figure}
        \centering
        \includegraphics[width=\hsize]{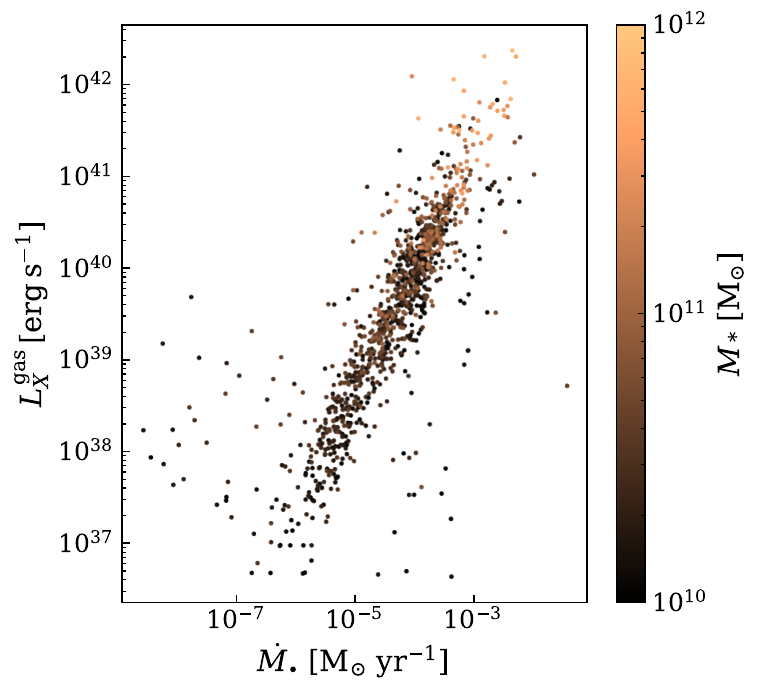}
        \caption{Gas luminosity within $\Rvir$ of the AGN cleaned sample as a function of the current accretion rate of the central SMBH $\dot{M}_{\bullet}$. The colorbar indicates stellar mass of the galaxy}
        \label{fig:lgas_mdot}
    \end{figure}

    The strong positive correlation between the slope of $S\beta$ galaxies and their total gas luminosity $L_{X,\rm{gas}}$ indicates that gas at the center of those galaxies is emitting more than in the outskirts. This suggests that a feedback mechanism is probably in place, injecting more energy into the central gas. One of the primary source of energy can be the activity of the SMBH at the center of each galaxy on which we will focus here.
    In order to quantify the central feedback in each galaxy in our AGN-cleaned sample, we show $L_{X,\rm{gas}}$ as a function of the current accretion rate of the central SMBH ($\dot{M}_{\bullet}$) in figure \ref{fig:lgas_mdot}. Each data point is colored by the stellar mass of the host galaxy. 
    We observe a tight coupling between the total luminosity of the gas component and the current accretion rate of the SMBH. This suggests that the main source of energy in our AGN-cleaned sample appears to be connected to the activity of the central SMBH at all stellar masses.
    Since we excluded overly X-ray bright AGN from this sample, this correlation is not driven by extreme cases and holds for faint AGNs as well.
    While the strong link between the SMBH and the regulation of the gas phase is expected \citep[e.g.][]{Hirschmann+2014, Steinborn+2015, Gaspari+2019, Truong+2020, Truong+2021a, Truong+2023}, high accretion rates generally also indicate a large gas reservoir in the vicinity of the SMBH.
    The results in figure \ref{fig:lgas_mdot} indicate that the injected energy in the innermost region heats up this gas quickly, so that this is immediately reflected in the gas luminosity.
    An additional mechanism injecting energy into the gas phase is associated to SN feedback. Especially in SF galaxies, SNII feedback can additionally heat up the gas and increase its luminosity and may trigger outflows which leave imprints in the CGM SB of SF galaxies \citep{Strickland+2004, Mineo+2012b, Li+2018}.
    
    While the $\beta$-profile is a simple and well established model, it is also assuming the gas to be in an isothermal state. Thus it can not account for the expected temperature gradients which are observed for galaxies and galaxy clusters alike \citep{Pratt+2007, Kim+2020}.
    Since the $\beta$-profile is a spherically symmetric profile, it can not account for asymmetries in the gas distribution either. In an idealized scenario for disk galaxies, distinct flow-patterns are expected to arise with outflows primarily happening perpendicular to the stellar disk and inflows circularizing at the edges of the stellar disk \citep[see e.g.][]{Tumlinson+2017, Stern+2023}.
    Recently, eROSITA revealed large-scale lobes perpendicular to the galactic disk of the MW \citep{Predehl+2020} which are compelling evidence for the aforementioned asymmetries. Simulations showed the presence of such asymmetries in mock X-ray images of simulated galaxies \citep{Truong+2021a, Truong+2023, ZuHone+2023}.
    However a recent study of edge-on star-forming galaxies in the Virgo galaxy cluster yielded only weak evidence for the presence of extraplanar hot gas \citep{Hou+2024} perpendicular to the stellar disc.
    We investigated our galaxy sample with respect to possible asymmetries by selecting disk-like SF galaxies. We binned them by their inclination angle with respect to the chosen l.o.s.\ (i.e.\ the z-axis of the simulation box.)
    We describe the exact setup in appendix~\ref{app:incl} and show the result in figure~\ref{fig:sb_incl}.
    We do not find any variation of the SB profiles with inclination angle in our sample. In low mass galaxies, the resolution limit of the simulation may smooth out emerging asymmetries.
    Moreover, the feedback from the central SMBH is distributed isotropically in its surroundings which further suppresses asymmetries. A spherically symmetric profile is thus a fair description in our case.
    
    We also verified the validity of the $\beta$-model description by reconstructing the X-ray emitting gas mass from the best-fit profiles of the simulated galaxies represented by a single $\beta$ model. The deprojected 3D density profile derived from the $\beta$-model takes the form 
    \begin{equation}
        \label{eq:beta_rho}
        \rho(r) = \rho_0 \lrb{1+\lrb{\dfrac{r}{r_c}}^2}^{-\frac{3\beta}{2}}\, ,
    \end{equation}
    where the central gas density is
    \begin{equation}
        \rho_0 = 2.21\mu m_p n_0 \, ,
    \end{equation}
    with
    $\mu$ being the mean molecular weight, $m_p$ the proton mass, and
    \begin{equation}
        \label{eq:n0}
        n_0 = \sqrt{\dfrac{S_0}{r_c \epsilon B(3\beta-0.5;0.5)}}\, .
    \end{equation}

\begin{figure}
        \centering
        \includegraphics[width=\hsize]{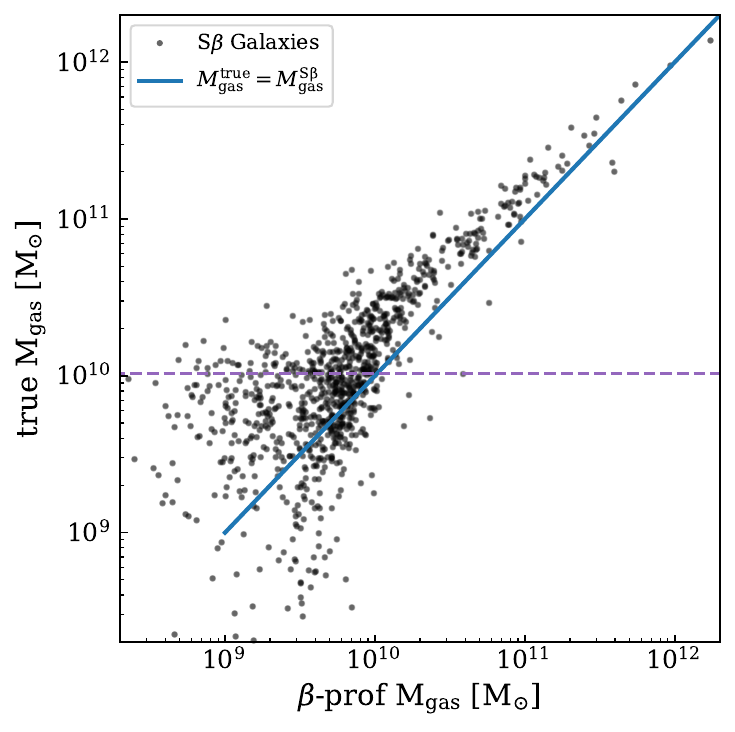}
        \caption{Comparison of the X-ray emitting gas mass from the simulation and the gas mass retrieved from single $\beta$-profiles using eq.\eqref{eq:beta_rho}-\eqref{eq:n0}. The solid blue diagonal line indicates equality between the two methods. The horizontal purple dashed line indicates $M_{\rm{gas}}^{\rm{true}}$ for a halo with 1000 gas resolution elements. Galaxies with $M_{\rm{gas}}^{\rm{true}}\lesssim 10^{10}\,\Msun$ have more poorly resolved atmospheres.
        }
        \label{fig:mgas_recstruct}
    \end{figure}

    In eq.~\eqref{eq:n0}, $n_0$ denotes the central number density at $r=0$, $\epsilon$ is the emissivity, $B(a,b)$ represents the validity of the beta function and $S_0$, $\beta$ and $r_c$ are the same as in eq. \eqref{eq:beta_prof} \citep[see also][for details]{Babyk+2018}. The resulting value
    For each galaxy we compute $\epsilon$ assuming an APEC model with the emissivity weighted average temperature and mass weighted metallicity of the X-ray emitting gas within $\Rvir$. The resulting total gas mass is then obtained from integration of eq. \eqref{eq:beta_rho} up to the virial radius.
    We show the comparison between mass of the total X-ray emitting gas in the simulation and the reconstructed gas mass from the beta profile in figure \ref{fig:mgas_recstruct}.
    We find good correspondence between intrinsic and reconstructed values for high gas masses ($M_{\rm{gas}}>10^{11}\,\Msun$), with some possible bias underestimating the reconstructed gas mass at intermediate mass ranges ($10^{10}\lesssim M_{\rm{gas}}\, [\Msun] \lesssim 10^{11})$.
    The scatter at very low masses ($M_{\rm{gas}}<10^{10}\,\Msun$) might be driven by a combination of the underlying simplifications as well as large uncertainties on the best-fit $r_c$ and $\beta$ values, originating from a poorly resolved atmosphere.

    Since the plasma in galaxies is generally cooler than in clusters, X-ray emission is mostly dominated by emission lines of various metal species. We therefore investigated possible trends between the slope of $S\beta$ galaxies and their mass weighted total metallicity.
    We did not find any correlation between $\beta$ and the metallicity within $0.1\Rvir$ or $\Rvir$, which means that the shape of our profiles is not determined by a metallicity gradient. Therefore the $\beta$-profile is an adequate description of the gas density in our case.

    \subsection{Contamination of the CGM emission}
    A major aspect when dealing with the SB of the CGM is the contamination by satellite galaxies \citep[see also][]{Zhang+2024a}. With their stellar mass and potentially high star-formation rates, satellite galaxies can host additional XRB sources. As indicated in figure \ref{fig:SB_mean}(b), not only does the gas component contribute towards the extended SB but XRB emission can have non-negligible contribution as well. 
    The show-case galaxy in figure \ref{fig:sid13633} illustrates how especially LMXBs can appear as a diffuse component in the outskirts of a galaxy.
    The exact contribution from XRBs is however highly dependent on the assumed spectral model \citep{Lehmer+2016,SVZ:I}. In our case, we use a relatively high column density when modeling the absorbed power law spectrum of XRBs compared to other studies \citep{Lehmer+2021, Riccio+2023, Kyritsis+2024}. Lower column densities would obviously yield higher contributions from XRBs in the soft band.
    Figure \ref{fig:sid13633} also shows the presence of actively accreting SMBH point sources which correlate with the position of the subhalos and would be likely masked in observations. 
    Since our AGN-exclusion criteria (section~\ref{sec:AGN-cleaning}) ensure
    that SMBHs associated to subhalos are less luminous than the central SMBH, we do not expect significant contamination.
    
    In order to explicitely mask radial bins where subhalos are dominant, we used a median filter on the gas SB profiles when fitting $\beta$-profiles and when constructing our mean profiles of figure \ref{fig:Compare}.
    While the median filter reliably detected substructure, the resulting masking of the affected radial bins rendered some of the SB profiles unusable. In those cases the profile was either dominated by substructures or had to few radial bins left and was consequently undetermined.
    
    Another aspect of XRB contamination comes from the association of XRBs with the main stellar body of a galaxy \citep[see][]{Grimm+2003, Gilfanov2004}. Our detailed model of XRB emission enabled us to quantify XRB contamination not only for the total luminosity (see figure \ref{fig:lx_mstar}) but also for the core-excised luminosity.
    Especially for low-mass galaxies, XRBs can have a significant contribution to the total luminosity of galaxies with low SB.

    In Fig. \ref{fig:lx_mstar} we showed that scaling relations between the halo mass $M_{500c}$ and both the total luminosity as well as the core-excised luminosity are in excellent agreement with observations. Especially in the group regime $M_{500c}>10^{13}\,\Msun$, where the gas component is most dominant, observed scaling relations are perfectly reproduced. 
    Since we include all major emitting components for the total luminosity $L_X$, we also verified that scaling relations are still consistent when only accounting for the gas component.
    We note that the largest luminosities at low halo masses are associated to galaxies with a bright AGN.
    
    \subsection{Scaling with global quantities}
    Both the QU and SF sample show distinct behavior at different radii with respect to variations in their global properties. In the QU case, all investigated properties mostly affect the normalization of the binned profiles. In the SF case normalization stays the same while the extent of the profiles change.
    We know from observations, that the hot atmospheres of elliptical galaxies are in hydro-static equilibrium with the gravitational potential such that global properties have a tight relation with the halo mass \citep{Kim+Fabbiano2015,Forbes+2017,Fabbiano2019}.
    In SF galaxies this relation is less clear and observations have shown that the total X-ray luminosity of SF galaxies is only weakly related to halo properties such as temperature \citep{Kim+Fabbiano2015}.
    Concerning $\fgas$ in our SF sample, we would have expected a stronger dependence on the normalization of the SB profiles. From the violin plots in figure \ref{fig:SB_prop} we can infer that the highest gas fraction bin is comprised preferentially of low mass galaxies for SF galaxies while there are preferentially more massive galaxies in the QU case.
    This would suggest that a high fraction of the hot gas is located in the halo outskirts of QU galaxies since the whole structure of the SB changes with $\fgas$. This is in line with typical formation  scenarios of QU galaxies where energetic feedback events from the central SMBH or major mergers redistributed and heated the gas.
    In turn, the hot gas appears to be more concentrated towards the center for our SF sample since the normalization of the SB profiles only changes for the inner $0.1\,\Rvir$ and is also supported by Fig.~\ref{fig:beta_corr}(c).
    This behavior is however not supported in the literature where shallower profiles are typically observed \citep{Bogdan+2013, Li+2017, Zhang+2024a}.
    The slope also increases for SF galaxies at low masses in our case where the resolution of the simulation may already be too low to properly resolve the hot gas atmosphere outside of $0.1\,\Rvir$.
    \subsection{Comparison to other simulations}
    Similar works on the X-ray SB in simulated galaxies have been conducted using different simulation suites.
    Notably, \citet{Oppenheimer+2020} analyzed full eROSTIA mock observations based on EAGLE \citep{Schaye+2015} and IllustrisTNG-100 (TNG100) \citep{Pillepich+2018} simulations. They made use of the pyXSIM package\footnote{http://hea-www.cfa.harvard.edu/~jzuhone/pyxsim/}, which is a python implementation of the \phox{} algorithm used in this work (see Sec.\ref{sec:Phox}).
    When generating the initial photon events from the simulation, they include a model for the MW foreground emission and absorption as well as a model for the background sources. However, they do not directly model the contribution from AGN and X-ray binaries.
    Additionally, they used SIXTE to account for instrumental effects on the resulting SB profiles. They split their galaxy sample in low ($M_*=10^{10.2-10.7}\,\Msun$) and high mass ($M_*=10^{10.7-11.2}\,\Msun$) star-forming and quiescent galaxies for both simulation sets and performed a stacking analysis on each respective subsample in a radial range of [10-300] kpc. 
    Their low-mass sample (see their Fig. 2, left) is therefore in a comparable MW mass range w.r.t. our analysis in Fig. \ref{fig:Compare}, on which we will focus here. 
    In general, they find flatter SB profiles for both simulations in the MW mass range compared to our findings, with EAGLE predicting less luminous star-forming galaxies than TNG100. Especially at (r $>$100 kpc) we predict an order of magnitude lower SB. At smaller radii our findings are consistent but interestingly show larger object by object variations, which we attribute to the differences in the details of the AGN feedback treatment in the simulations.
    Both EAGLE and TNG100 confirm the significant increase in central SB for star-forming galaxies compared to quiescent galaxies which we observe in our simulations.
    Furthermore, \citet{Truong+2020} found that blue galaxies in TNG are an order of magnitude brighter in X-ray than red galaxies at fixed stellar mass, confirming this dichotomy.
    For radii $r>50\,\rm{kpc}$, we find that SF and QU galaxies have similar SB profiles which is consistent with EAGLE galaxies in \citet{Oppenheimer+2020}, whereas TNG100 predicts larger SB for star-forming galaxies at all radii.
    We note that \citet{Oppenheimer+2020} considered the X-ray photon emission from the gas included within a sphere of $3\cdot R_{200}$ around each galaxy, which is significantly larger than the volume inspected here (see Sec. \ref{sec:data_set}). Some contamination in the outskirts, due to gas in the galaxy surrounding not associated to the galaxy itself, can thus be present and contribute to the flattening of the SB profiles. 
    Moreover, those authors performed full mock observations aiming to test CGM detectability, while we focused on intrinsic emission and its connection to global properties of our galaxies. We thus expect some differences introduced by the instrumental effects, such as in the central SB due to effects from the eROSITA PSF or in the galaxy outskirts due to the treatment of the background in observations and full mocks.

\section{Summary} \label{sec:Conclusions}
In this study, we present results on the X-ray SB of simulated galaxies from the \textit{Magneticum Pathfinder} set of simulations. We made use of the virtual X-ray photon simulator \phox{} \citep{Biffi+2012, Biffi+2018agn, SVZ:I} to produce highly sophisticated spectral models of individual galaxies where we properly account for the multi-temperature and metallicity distribution of the gaseous component. The emission from SMBH sources and XRBs can be accounted for individually and self-consistently without the need for empirical modeling of their contribution. We accounted for an AGN population in our galaxy sample by applying exclusion criteria motivated from the literature. We focused our analysis on an AGN-cleaned sample uncovering the following aspects:

\begin{itemize}
    \item We determined mean and median SB profiles of normal star-forming ($\log(\rm{sSFR})>-11$) and quiescent galaxies and quantified the contribution of different components. We find that SF galaxies have elevated total SB in their central regions compared to QU galaxies up to a scale-free radius of $0.1\,\Rvir$ and comparable SB in the outskirts. The average contribution from XRBs towards the total SB is between 30\%-50\% in the inner $0.1\,\Rvir$ and $\gtrsim 10\%$ for larger radii, which we attribute to the presence of substructures. 

    \item We compared median SB profiles of the gas component in
    a $M_*$-matched subsample of our galaxies to recent observational results from the eROSITA collaboration, where they obtained CGM SB from stacking MW-mass galaxies. We are in good agreement with results form \citet{Zhang+2024a} who have a similar median redshift distribution in their sample compared to our fixed redshift. We find no significant difference between mean CGM profiles of SF and QU galaxies in our sample for the MW mass range ($M_*=10^{10.5-11}$). Results from \citet{Comparat+2022} are in disagreement with our sample which may be due to instrumental effects.
    
    \item We computed hot gas fractions and temperatures of each galaxy and found positive correlations between the extent and normalization of SB profiles and the respective property. Correlations are generally stronger for QU galaxies than for SF galaxies. All our galaxies have $\fgas<f_{\rm{bary}}$ accounting for hot gas within $\Rvir$. We compared the SB of our sample with SB of observed galaxies for which the same properties ($M_*$, $\fgas$, $k_B T$) were available and found good agreement for SF galaxies. We found lower normalization for our QU sample compared to the selected observations. The selected observations are however more consistent with our excluded AGN sample. We attribute this fact to the observational sample including more BCG galaxies which makes them brighter than typical QU galaxies.
    
    \item We fitted $\beta$-models to SB profiles of the gas component of our AGN-cleaned galaxy sample using a least-square algorithm. We found that most of our sample is best reproduced by a single slope $\beta$-profile. We compared the profile slope to intrinsic properties for each galaxy and found that SF galaxies have consistently steeper profiles compared to QU galaxies. We found strong correlation between steepness and total gas luminosity, in agreement with observational results. We found large scatter in the slope of low $M_*$ galaxies in both the QU and SF sample. The slope of the SB profiles seems uncorrelated with $\fgas$ and $T_{\rm{gas}}$ for SF galaxies and weakly correlated for QU galaxies while being consistent with observational results.
    
    \item We compared the $M_{500c}-L_{X}$ relation of our galaxy sample to stacking results of A15 and Z24b for galaxy-mass halos and also compared to the scaling relations for galaxy groups from \citet{Lovisari+2015}. While the intrinsic scatter in our sample increases for low halo masses we found excellent agreement with observational results at all halo masses. The scatter is a result of subgrid physics in the underlying cosmological simulation. 
    The estimated median XRB contribution is consistent with theoretical models and and can account for the total luminosity of galaxies with the lowest SB. 
    
    \item Additionally, we compared the CGM luminosity of our galaxies by using an annulus similar to A15 ($r>0.15\,R_{500c}$) and found median luminosities in excellent agreement with their stacking results. At low halo masses, LMXBs can contribute significantly to the overall CGM luminosity which may be due to satellite galaxies or a diffuse stellar component outside of the central galaxy.

    \item We find that the main engine for the gas luminosity and consequently the increase in steepness in our galaxy sample is the central SMBH of each galaxy. We find a strong dependence of the global gas luminosity on the current accretion rate of the central SMBH, in the whole range of host galaxy stellar masses investigated. While the SMBH injects feedback energy isotropically into the surrounding in the simulation, the internal energy of neighboring gas is increased. 
    A higher $\dot{M}_{\bullet}$ generally also suggests that there is more gas in the vicinity of the SMBH in order to sustain the accretion. The energy is then released close to the SMBH which leads to a centrally localized increase in luminosity and consequently to a steeper SB profile.
\end{itemize}

Ultimately, we were able to show that the CGM of galaxies can be resolved in modern cosmological simulations and can offer a unique way of benchmarking observational results. Empirically well studied relations for the X-ray emission of galaxies arise from the simulation self-consistently and provide predictions for the distribution of the gaseous halo out to the virial radius.
However, it is challenging to directly observe the hot CGM in emission due to the emissivity being proportional to the density squared. The detection of the CGM out to large radii has been confirmed in emission using stacking procedures. While stacking can retrieve average properties of the underlying sample, information about the diversity of trends and scatter in those properties, such as metallicity, temperature or gas fraction, is lost.
Our study can be therefore useful to better understand the underlying properties of galaxies and their systematics in commonly used observational techniques.
In the future, current and proposed X-ray missions using calorimetry-based detectors, such as XRISM \citep{XRISM} or LEM \citep{LEM}, may provide additional insights by directly measuring gas emission lines and setting more stringent constraints on feedback models and enrichment in the CGM.

\begin{acknowledgements}
SVZ and VB acknowledge support by the \emph{Deut\-sche For\-schungs\-ge\-mein\-schaft, DFG\/} project nr. 415510302. KD acknowledges support through the COMPLEX project from the European Research Council (ERC) under the European Union’s Horizon 2020 research and innovation program grant agreement ERC-2019-AdG 882679.
This research was supported by the Excellence Cluster ORIGINS which is funded by the Deutsche Forschungsgemeinschaft (DFG, German Research Foundation) under Germany's Excellence Strategy – EXC-2094 – 390783311.
The calculations for the hydrodynamical simulations were carried out at the Leibniz Supercomputer Center (LRZ) under the project pr83li. We are especially grateful for the support by M. Petkova through the Computational Center for Particle and Astrophysics (C2PAP). SVZ would like to thank IK and IM for helpful comments and insightful discussions.  
\end{acknowledgements}

%
%
\bibliographystyle{aa}
\bibliography{references}

\begin{appendix}

\section{Behavior of the AGN sample}\label{app:AGN_sample}
\begin{figure*}
        \centering
        \includegraphics[width=\hsize]{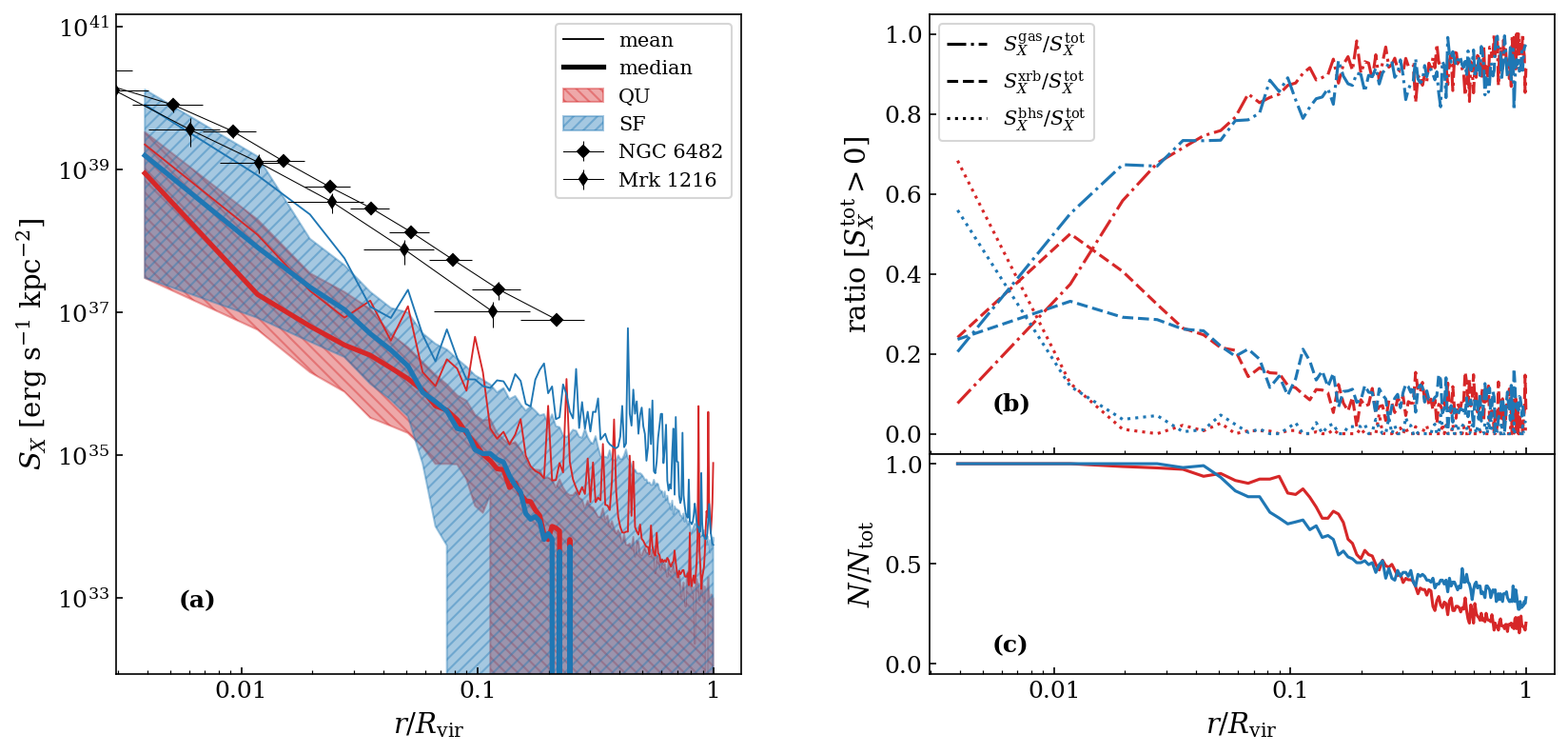}
        \caption{
            Same as figure \ref{fig:SB_mean} for the excluded AGN sample. Additionally, we include SB profiles from BCG like galaxies NGC 6482 and Mrk 1216.
        }
        \label{fig:SB_mean_AGN}
    \end{figure*}

In Fig.~\ref{fig:SB_mean_AGN} we present the mean and median SB profiles of all galaxies excluded by our exclusion criteria from Sec. \ref{sec:AGN-cleaning}. The excluded sample mostly comprises X-ray AGN-dominated galaxies as well as 6 brightest group galaxies (BGGs) with a stellar mass $M_*>10^{12}\,\Msun$.
Compared to the cleaned sample, the excluded sample is more than an order of magnitude brighter in the center, which is due to the X-ray bright AGN emission. This can be seen in Fig.~\ref{fig:SB_mean_AGN}(b) where emission of the central SMBH contributes more than 60\% towards the total SB in the center.
Furthermore, the difference between SF (blue) and QU (red) galaxies in the central $0.1\,\Rvir$ is weaker in the excluded sample than in the cleaned sample.
At large radii, profiles are similar to those reported in Fig.~\ref{fig:SB_mean} for normal galaxies. However, the excluded sample shows slightly more extended emission for SF galaxies. This is caused by the BGGs in the excluded sample which are also highly star-forming.
In Fig.~\ref{fig:SB_mean_AGN} we also include the SB profiles of BCG-like galaxies NGC 6482 \citep{Buote+2017} and Mrk 1216 \citep{Buote+2018}, which are in better agreement with the excluded sample compared to the cleaned sample.

\section{Examples}\label{app:examples}
\begin{figure*}
    \centering
    \includegraphics[width=.49\linewidth]{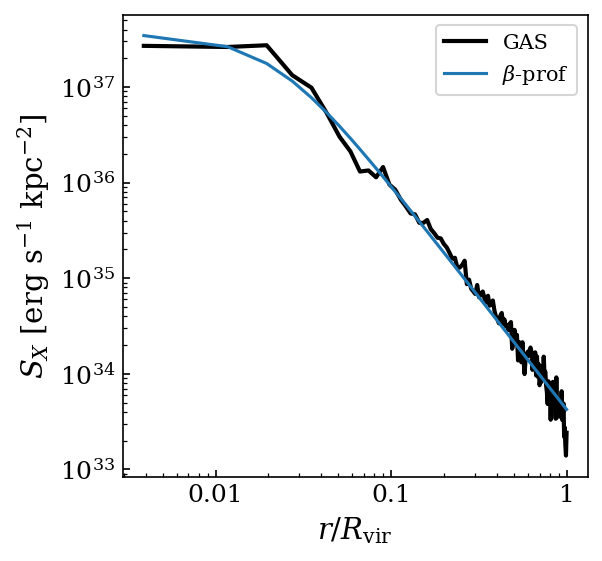}
    \includegraphics[width=.49\linewidth]{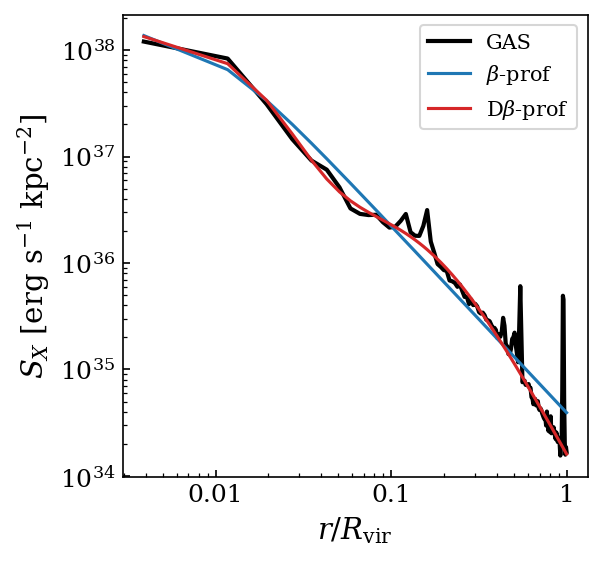}
    \caption{Exemplary SB profiles (\textit{solid black}) from the S$\beta$ category (\textit{left}) and D$\beta$ category (\textit{right}). The solid blue line shows the best fit single $\beta$-profile and the solid red line shows the best-fit double $\beta$-profile. }
    \label{fig:beta_example}
\end{figure*}

In Fig. \ref{fig:beta_example} we show two exemplary SB profiles (black) from our sample. The left panel shows a galaxy from the $S\beta$ category which is best represented by a single $\beta$-profile (blue).
The right panel shows a galaxy from the $D\beta$ category which is best represented by a double $\beta$-profile (red). For reference, we also included the best fit single $\beta$ profile in the right panel to illustrate the difference.
We note, that the $D\beta$ example also has some substructure within $\Rvir$ which is seen as large jumps in the SB profile. Those were masked using a median filter before fitting the profile.

\section{Galaxy main sequence}
\label{app:mainseq}
\begin{figure}
        \centering
        \includegraphics[width=0.9\hsize]{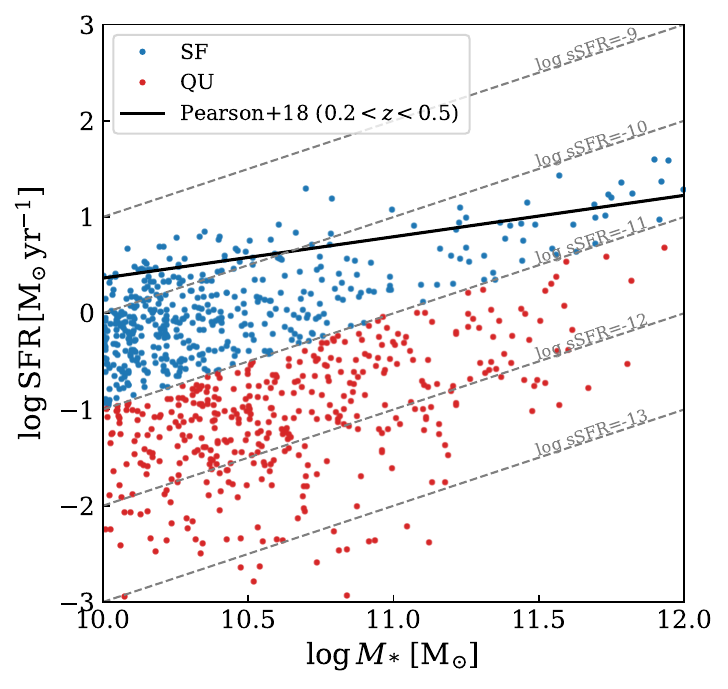}
        \caption{
            The main sequence of our total galaxy sample extracted from the simulations. The dashed diagonal lines indicate constant specific star-formation rates in $\rm{Gyr^{-1}}$. The solid black line indicate the empirical mainsequence from \citep{Pearson+2018} for the redshift range $z=0.2-0.5$. Blue colors are SF galaxies and red are QU galaxies which we distinguish using the $\log \rm{sSFR\,[\rm{yr}^{-1}]}=-11$
        }
        \label{fig:mainseq}
    \end{figure}

In Fig. \ref{fig:mainseq} we show the SFR-$M_*$ relation for our galaxy sample, color-coded by SF (blue) and QU (red) galaxies. The black solid line indicates the observed main sequence from \citet{Pearson+2018} in the redshift range $0.2<z<0.5$. The gray dashed diagonal lines indicate the relation for constant values of sSFR. 

We note that low-mass SF galaxies lie below the main sequence by 0.5 dex, indicating that they typically have less star-formation than expected. 

\section{Inclination}
\label{app:incl}
\begin{figure}
        \centering
        \includegraphics[width=\hsize]{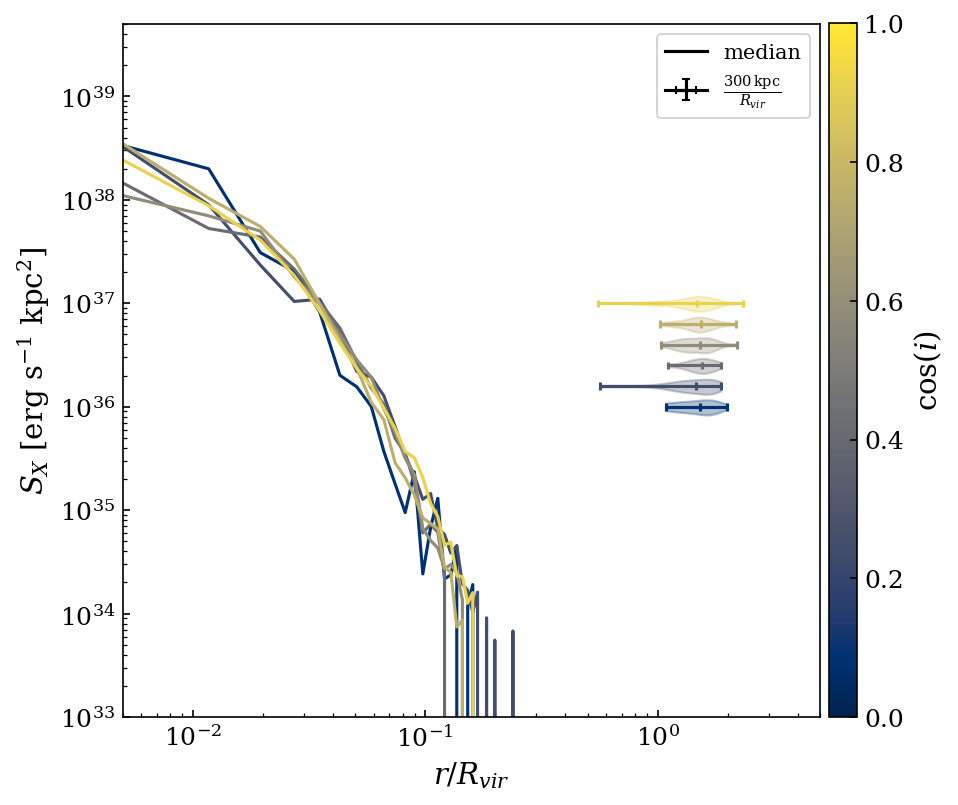}
        \caption{
            Similar to figure \ref{fig:SB_prop}. Colors indicate the cosine of the inclination angle of star-forming disk galaxies within each bin.
        }
        \label{fig:sb_incl}
    \end{figure}

In order to quantify the imprints of galaxy orientation with respect to the l.o.s., we investigate the inclination dependence of SB profiles in star-forming disk galaxies.
In our framework, the inclination angle $i$ of each galaxy can be computed from the scalar product between the fiducial l.o.s.\ ($\pmb{\hat{e}_z}$) and the specific angular momentum of the stellar component ($\pmb{j_*}$), such that
$$\cos i = \dfrac{\pmb{\hat{e}_z}\cdot\pmb{j_*}}{\lVert j_* \rVert}\,.$$
The galaxy would be seen edge-on for $\cos i = 0$ and face-on for $\cos i = 1$. 
For this test, we only consider star-forming galaxies with a $b$-value $b>-4.35$, where 
$$b=\log \lrb{\dfrac{j_*}{\rm{kpc\,km\,s^{-1}}}} - \dfrac{2}{3}\log\lrb{\dfrac{M_*}{\Msun}}\,, $$
is a measure of the galaxy morphology \citep{Teklu+2015} and the numerical value was chosen according to \citet{Schulze+2020} for disk galaxies.
This constrains our sample to only include truly disk-like galaxies, allowing for a meaningful interpretation of the inclination angle.
In figure \ref{fig:sb_incl} we show mean SB profiles of our disk-like subsample, where galaxies were binned according to the value of $\cos i$ (with colors indicating the bin centers). Colored violins have the same meaning as in Fig. \ref{fig:SB_prop}.
We do not find any strong dependence of the SB profile on galaxy inclination angle. 
From a theoretical perspective, it is intuitive to assume preferential outflow directions perpendicular to the galactic disc due to the path of least resistance.
Numerical studies using different simulations find in fact asymmetric outflow patterns in X-ray mock observations of disc galaxies \citet{Truong+2021b, Schellenberger+2023, Truong+2023}.
Observationally, studies of X-ray emission around disk galaxies did not find however any enhanced signal perpendicular to the disc \citep{Bogdan+2013, Li+2018, Hou+2024}.

\end{appendix}

\end{document}